\documentclass[twocolumn,showpacs,preprintnumbers,amsmath,amssymb]{revtex4}

\usepackage{graphicx}
\usepackage{dcolumn}
\usepackage{bm}

\def\mr#1{\mathrm{#1}}

\newcounter{ichi}
\newcounter{ni}
\newcounter{san}
\newcounter{yon}
\newcounter{go}
\newcounter{roku}
\makeatletter
\@addtoreset{footnote}{page}
\makeatother
 
\begin{document}

\preprint{Accepted for publication in PRD}

\title{High-energy cosmic-ray nuclei from high- and low-luminosity\\ 
gamma-ray bursts and implications for multi-messenger astronomy}

\author{Kohta Murase$^{1}$}
 \email{kmurase@yukawa.kyoto-u.ac.jp}%
\author{Kunihito Ioka$^{2}$}%
\author{Shigehiro Nagataki$^{1}$}%
\author{Takashi Nakamura$^{3}$}%
\affiliation{%
$^{1}$Yukawa Institute for Theoretical Physics, Kyoto University,\\
Oiwake-cho, Kitashirakawa, Sakyo-ku, Kyoto, 606-8502, Japan\\
$^{2}$Theory Division, KEK (High Energy Accelerator Research Organization),
1-1 Oho, Tsukuba 305-0801, Japan\\
$^{3}$Department of Physics, Kyoto University, Kyoto 606-8502, Japan
}%

\date{May 2}
                        
\begin{abstract} 
Gamma-ray bursts (GRBs) are one of the candidates of
ultra-high-energy ($\gtrsim {10}^{18.5}$ eV) cosmic-ray (UHECR) sources.
We investigate high-energy cosmic-ray acceleration including heavy
nuclei in GRBs by using Geant 4, and discuss its various implications, 
taking both of high-luminosity (HL) and low-luminosity (LL)
GRBs into account. This is because LL GRBs may also make a significant 
contribution to the observed UHECR flux if they form a distinct population.
We show that not only protons but also heavier nuclei can be accelerated up 
to ultra-high energies in the internal, (external) reverse and forward shock
models. We also show that the condition for ultra-high-energy heavy
nuclei such as iron to survive is almost the same as that for $\sim$ TeV 
gamma-rays to escape from the source and for high-energy neutrinos
not to be much produced. The multi-messenger astronomy by
neutrino and GeV-TeV gamma-ray telescopes such as IceCube and KM3Net, GLAST and
MAGIC will be important to see whether GRBs can be accelerators of 
ultra-high-energy heavy nuclei. We also demonstrate expected spectra
of high-energy neutrinos and gamma rays, and discuss their detectabilities.
In addition, we discuss implictaions of the GRB-UHECR
hypothesis. We point out, since the number densities of HL-GRBs and LL-GRBs are
quite different, its detemination by UHECR observations is also important.
\end{abstract}

\pacs{95.85.Ry, 98.70.Rz, 25.20.-x, 14.60.Lm, 96.50.Pw, 98.70.Sa}
\maketitle

\section{\label{sec:level1}Introduction}
Gamma-ray bursts (GRBs) and supernovae (SNe) are the most powerful
phenomena in the universe. The latter is believed to be the origin of 
high-energy cosmic rays below the knee $\approx {10}^{15.5}$
eV. The former could also accelerate baryons up to high energies if
the dissipation process is due to shock dissipation.
Waxman, Milgrom \& Usov and Vietri suggested that ultra-high-energy
cosmic rays (UHECRs) are produced in GRBs \cite{Wax1,Mil1,Vie1}. 
Their suggestions were based on the two arguments. First is the 
requirement that relativistic outflows that make GRBs satisfy various 
conditions for baryons to be accelerated up to greater than ${10}^{20}$ eV.
Second is that the energy generation rate required to account for the
observed UHECR flux is comparable to the energy generation rate of 
observed gamma rays. The latter argument depends on the local GRB rate 
which is not well known observationally. 
If the local cosmological high-luminosity GRB (HL GRB) rate is not
high enough, which may be suggested by recent observations
\cite{Le1,Gue1,Kis1}, the required baryon loading becomes larger 
\cite{Wic1,Mur1,Ber1}. In addition, if the UHECR spectrum at the
source is steeper than that with the spectral index $p \sim 2.0-2.3$ (as
expected in the dip model, $p \sim 2.4-2.7$ \cite{Ber1,Alo1}), the 
total nonthermal
cosmic-ray energy of GRBs, necessary for explaining the observed
UHECR flux, would be larger than the radiation energy. Despite
such caveats, the GRB-UHECR hypothesis is still one of the most
interesting possibilities for explanation of observed UHECRs, and it 
should be examined by future observations.  

Recent observations have tentatively suggested that 
some GRBs may form a different class (see, e.g.,
\cite{Lia1,Gue2}). 
GRBs in this class are called 
low-luminosity GRBs (LL GRBs) or sub-energetic GRBs, which may be 
more common than HL GRBs. If this speculation is true, they can 
provide the significant energy as high-energy cosmic rays, neutrinos
and gamma rays. Murase et al. \cite{Mur2} suggested 
that cosmic rays can be accelerated up to ultra-high energies
($\gtrsim {10}^{18.5}$ eV) in LL GRBs, and LL GRBs may make an important 
contribution to the observed high-energy cosmic-ray flux and the 
diffuse neutrino background \cite{Mur2,Gup1}. Such LL GRBs were
associated with energetic supernovae called hypernovae. As in cases of 
supernovae, cosmic-ray acceleration is expected at shocks formed by the
high-velocity ejecta. In fact, if the magnetic field is sufficiently 
amplified, cosmic-rays can be accelerated up to very high energies,
even to ultra-high energies. Wang et al. \cite{Wan1,Wan2} suggested that 
they could explain observed high-energy cosmic rays above the second 
knee, under the assumption that LL GRBs have the enough local rate higher 
than that of HL GRBs. 

Recently UHECR observations have been greatly advanced thanks to 
(Sourth) Pierre Auger Observatory (PAO). 
The UHECR spectrum deduced by PAO suggests the existence of 
the Greisen-Zatsepin-Kuz'min (GZK) cutoff energy which
arises from the photomeson production process between UHECRs
and cosmic microwave background (CMB) photons \cite{Gre1,Zat1}.  
In addition, PAO recently revealed the 
anisotropy and rejected the isotropy of UHECRs above 
60 EeV at 3 $\sigma$ level \cite{PAO1,PAO2}. 
It suggests that the 
distribution of UHECRs traces the matter 
distribution in the universe, although the origin of UHECRs is still unknown,
that is, AGNs, GRBs and other possibilities remain (see the latest
review for AGN and GRBs, e.g., \cite{Der1}).  
Interestingly, the preliminary elongation rate data by PAO showed the
break around $\sim {10}^{18.35}$ eV \cite{PAO3}. Whether heavier
nuclei becomes 
more important above this break as the energy increases (which seems 
conflict with HiRes data) is still under debate. It is because there are 
large uncertainties in hadronic interactions. 
On the other hand, recent 
results on UHECRs above 60 EeV by PAO may infer that protons are 
dominant sources \cite{PAO1,PAO2}, although some authors dispute 
\cite{Gor1,Wib1}. Although the issue of the composition of UHECRs requires 
more and more studies, present data seem to allow the 
presence of significant fractions of ultra-high-energy (UHE) nuclei 
around the GZK cutoff energy, even if protons are dominant. Therefore, 
it is one of the important issues whether UHE nuclei can survive
(i.e., UHE nuclei are not photodisintegrated and not depleted due to 
photomeson production during the dynamical time) in UHECR production sites.

In this paper, we study acceleration of high-energy cosmic-ray nuclei
in very detail by exploiting numerical calculations, and discuss
various implications for multi-messenger astronomy.  
In Sec. \Roman{ni}, we show that cosmic rays (protons and heavier
nuclei) can be accelerated up to ultra-high energies in both of HL
GRBs and LL GRBs. We consider the internal, (external) 
reverse and forward shock models.
 
If cosmic rays can be accelerated up to sufficiently high energies,
high-energy neutrino production via the photomeson production process 
and/or $pp$ reaction is unavoidable. Now, large neutrino detectors
such as IceCube \cite{Ahr1}, KM3Net
\cite{Kat1} are being constructed. ANITA \cite{Bar1} and PAO
\cite{Van1} can also detect very high-energy neutrinos. We have
chances to see such high-energy neutrino signals, which are important as
a probe of cosmic-ray acceleration, in the future.
High-energy neutrino signals from HL GRBs were predicted in the context of the 
standard scenario of GRBs assuming that observed UHECRs come from 
GRBs \cite{Wax2,Wax3,Dai1,Der2}. We also investigated the neutrino 
emission from GRBs in very detail \cite{Mur1,Mur2,Mur3,Mur4}, and
various predictions in the \textit{Swift} era are summarized in Ref. 
\cite{Mur4}. In Sec. \Roman{san}, we demonstrate predicted neutrino
spectra in some of our models and discuss implications for neutrino astronomy. 
We can see that neutrino signals highly suppressed when very
heavy nuclei like iron can survive. 

In Sec. \Roman{yon}, we also demonstrate gamma-ray spectra
for some models when heavy nuclei can be accelerated up to ultra 
high energies and survive. In this section, we also show that very 
high-energy gamma rays with GeV-TeV energies can escape from 
the source under conditions where UHE heavy nuclei such as iron can
survive. We can expect to detect high-energy gamma-ray signals
by upcoming Gamma-Ray Large Area Telescope (GLAST) and Cherenkov
telescopes such as MAGIC as long as the source is nearby. Otherwise, 
high-energy gamma rays above $\sim$TeV are attenuated by the cosmic 
microwave/infrared background (CMB/CIB) photons (\cite{Mur6} and 
references there in).

In Appendix F, We also discuss various implications of the
GRB-UHECR hypothesis (including the HLGRB-UHECR, LLGRB-UHECR and 
Hypernova-UHECR hypotheses) for UHECR astronomy. We see that 
the determination of the source number density of UHECR sources and
extragalactic magnetic field (EGMF) strength is essential to test the 
hypothesis. In this paper, we shall hereafter use notations such 
as $Q_x=Q/10^x$ in cgs units.

\section{Acceleration of Cosmic-Ray Nuclei in GRBs}
In the standard model of GRBs (for reviews, see, e.g.,
\cite{Pir1,Mes1,Zha1}), both of the prompt and afterglow
emission are attributed to electromagnetic radiation from relativistic 
electrons accelerated in shocks. If relativistic outflows that make
GRBs contain baryons, not only electrons but also protons (and heavier
nuclei) will be accelerated. Although the detail of the acceleration
mechanism is poorly known, we usually assume that the first-order
Fermi acceleration mechanism works in GRBs, where the 
distribution of nonthermal cosmic rays is given by a power-law 
spectrum under the test-particle approximation (see reviews, e.g., 
\cite{Bla1}). 
Therefore, we have \cite{Mur4,Typ1} 
\begin{equation}
\frac{d n_{\rm{CR}}}{d {\varepsilon}_{\rm{CR}}} = \frac{U_{{\rm{CR}}}}
{\int_{{\varepsilon}_{\rm{CR}}^{\mr{min}}}^{{\varepsilon}_{\rm{CR}}
^{\mr{max}}}  d \varepsilon_{\rm{CR}} {\varepsilon_{\rm{CR}}}^{-(p-1)}}
\varepsilon_{\rm{CR}}^{-p},
\end{equation}
where $\varepsilon_{\rm{CR}}$ is the energy of cosmic rays in the 
comoving frame, $U_{\rm{CR}}$ is the comoving cosmic-ray energy 
density, and $p$ is the spectral index. As a shock becomes highly 
relativistic, the spectral index can deviate from values for
non-relativistic shocks. 
In the diffusive small pitch-angle scattering regime, $p \simeq 2.2$ is 
obtained in the ultra-relativistic limit \cite{Kes1}. But the
large-angle scattering can lead to harder spectral indices
\cite{Vie2,Aoi1}. For simplicity, we shall use $p=2$ which is often
expected for non-relativistic or mildly relativistic shocks with 
the compression ratio $r_{\rm{c}}=4$. 
The acceleration time scale is written as $t_{\rm{acc}} = (\eta
\varepsilon_N)/(ZeBc)$, which could be applied to the second-order Fermi 
acceleration mechanism when the Alfv\'en-wave speed is close to the
light speed \cite{Rac1,Gal1,Ach1}. 
Here, $\varepsilon_N$ is the cosmic-ray energy in the comoving frame
and $\eta$ is a pre-factor. For the most efficient acceleration, we
can expect $\eta \sim (1-10)$ \cite{Rac1}. Throughout this paper $\eta =1$
is used to obtain the \textit{upper} limit of the maximum energy. Note
that $\eta \sim 10$ would be a more realistic value, and the
acceleration time scale with $\eta = 10$ is demonstrated in Ref. \cite{Mur1}. 

The maximum energy of cosmic-rays is determined by several criteria. 
One is derived from the requirement that the Larmor radius of a
particle should be smaller than the effective size of the acceleration 
region (the Hillas condition). In addition, the acceleration time scale 
should be smaller than the dynamical time scale $t_{\rm{dyn}}$ (which
is essentially the same as the Hillas condition in our interested cases)
and escape time scale $t_{\rm{esc}}$. The maximum energy is also limited by 
the total cooling time scale $t_{\rm{cool}}^{-1} \equiv t_{N\gamma}^{-1} + 
t_{\mr{syn}}^{-1} + t_{\mr{IC}}^{-1} + t_{\mr{ad}}^{-1}$. Here,
$t_{\mr{syn}}$ is the synchrotron cooling time scale, $t_{\mr{IC}}$ is
the inverse-Compton cooling time scale and $t_{\mr{ad}}$ is the
adiabatic cooling time scale. In cases of GRBs, we shall use
$t_{\rm{ad}} \approx t_{\rm{dyn}}$ \cite{Wax2,Mur4,Der4,Com0}. 
$t_{N \gamma}$ is the photohadronic time scale which includes 
the photodisintegration, photomeson production and photopair
production processes. 
For protons with sufficiently high energies, the photomeson production
process is the most important, whose energy loss time scale is given by
\begin{equation}
t^{-1}_{p\gamma}(\varepsilon _{p}) = \frac{c}{2{\gamma}^{2}_{p}} 
\int_{\bar{\varepsilon}_{\mr{th}}}^{\infty} \! \! \! d\bar{\varepsilon} \, 
{\sigma}_{p\gamma}(\bar{\varepsilon}) {\kappa}_{p}(\bar{\varepsilon})
\bar{\varepsilon} \int_{\bar{\varepsilon}/2{\gamma}_{p}}^{\infty} 
\! \! \! \! \! \! \! \! \! d \varepsilon \, {\varepsilon}^{-2} 
\frac{dn}{d\varepsilon}, \label{tpg}
\end{equation}
where $\bar{\varepsilon}$ is the photon energy in the rest frame of
proton, $\gamma _{p}$ is the proton's Lorentz factor, $\kappa _{p}$ is
the inelasticity of proton, and $\bar{\varepsilon} _{\mr{th}} \approx
145$ MeV is the threshold photon energy for photomeson production. 
For heavier nuclei than proton, both of the photodisintegration and photomeson
production processes are important, whose time scales are given by 
the similar expression to Eq. (\ref{tpg}) \cite{Ste1}.
In order to decide whether a kind of UHE nuclei can survive or not,
we calculate the interaction time scale of photodisintegration and photomeson
production by using the numerical simulation kit Geant4 \cite{Ago1},
which includes the cross section data based on experimental data \cite{PDG1}.
As seen later, it is important to use the accurate cross section in
the high-energy range. 
Although we evaluate $t_{N \gamma}$ by numerical calculations, 
the simple analytic treatment is often
useful. The most frequently used approximation for the photomeson 
production process is the $\Delta-$resonance 
approximation (see, e.g., \cite{Wax2,Mur3}). The corresponding one for 
the photodisintegration process is the Giant-Dipole-Resonance (GDR) 
approximation. 
Similarly, we can apply it to the GDR approximation for a
broken power-law photon spectrum. We have
\begin{eqnarray}
t_{N\gamma}^{-1} \simeq \frac{U_{\gamma}}{5 \varepsilon ^b}
\!\!\! &c& \!\!\! \sigma _{\rm{res}} 
\frac{\Delta \bar{\varepsilon}}{\bar{\varepsilon} _{\rm{res}}}
 \left\{ \begin{array}{rl} 
{(E_N/E_N^b)}^{\beta-1}\\
{(E_N/E_N^{b})}^{\alpha-1} 
\end{array} \right. , \label{tNg2}
\end{eqnarray}
where $E_{N}^{b} \simeq 0.5 \bar{\varepsilon}
_{\rm{res}}m_N c^2 \Gamma ^2/{\varepsilon}_{\rm{ob}}^{b}$, 
$\alpha$ is a photon index in lower energies while $\beta$ in
higher energies. The parameter regions for the upper and lower columns 
are $E_N < E_N^b$ and $E_N \geq E_N^b$, respectively. Here 
$\varepsilon_{\rm{ob}}^b$ is the break energy
measured by the observer in the local rest frame, and $U_{\gamma}$ is
the total photon energy density. For example, a Band function which
reproduces spectra of the prompt emission, can be approximated by a
broken power-law spectrum. Spectra of afterglows are also expressed by
several segments of power-law spectra.   
The cross section at the resonant energy 
$\bar{\varepsilon}_{\rm{res}}$ is expressed as $\sigma_{\rm{res}}$.
For GDR of nucleon, we use $\sigma _{\rm{GDR}} \sim 1.45 \times 
{10}^{-27} \, {\mr{cm}}^2 \, A$, $\bar{\varepsilon}
_{\rm{GDR}} \sim 42.65 A^{-0.21}$ MeV and $\Delta \bar{\varepsilon} \sim 8$  
MeV \cite{Ste1,Pug1,Kar1,All1}. In the case of $\Delta-$resonance of proton, 
we use $\sigma _{\Delta} \sim 4 \times {10}^{-28} \, {\mr{cm}}^2$, 
$\bar{\varepsilon}_{\Delta} \sim 0.3$ GeV, 
$\Delta \bar{\varepsilon} \sim 0.2$ GeV and 
${\kappa}_p \sim 0.2$ \cite{Wax2,Mur3}.
These approximations reproduce numerical results well except in
high energies. In high energies, effects of 
non-GDR and/or non-$\Delta$-resonance such as fragmentation and
multi-pion production can become important, as seen later. This
tendency can also be seen in $t_{p \gamma}$. As pointed out in 
Ref. \cite{Mur1}, the effect of non-$\Delta$-resonance such as multi-pion
production becomes moderately important in high energies for 
spectra of the prompt emission where $\alpha \sim 1$.

For later discussions, let us introduce optical depths for
photodisintegration of nuclei and photomeson production of protons,
and define $f_{N \gamma} \equiv t_{\rm{dyn}}
/t_{N \gamma}$ and $f_{p \gamma} \equiv t_{\rm{dyn}}
/t_{p \gamma}$, respectively. These quantities express whether cosmic rays can
survive in the source or not. For example, cosmic rays can survive 
from the photodisintegration and photomeson production processes 
if $f_{N \gamma}<1$, while not if $f_{N \gamma} \geq 1$.  

In the following subsections, we shall show that high-energy
cosmic-ray nuclei can be produced in the internal and external shock
models for HL and LL GRBs. GRBs may be even UHECR sources. In such
cases, note that cosmic rays have to be accelerated above ${10}^{20}$
eV from observations of the highest energy events and the
existence of the bump around the GZK cutoff energy. We shall allow for 
possibilities that observed highest UHECRs are UHE nuclei. It is
because, although the arrival distribution and spectrum of UHECRs may
be consistent with the proton model, some authors claimed that UHE nuclei 
are more important. 

\subsection{Internal Shock Model}
\begin{figure}[t]
\includegraphics[width=0.97\linewidth]{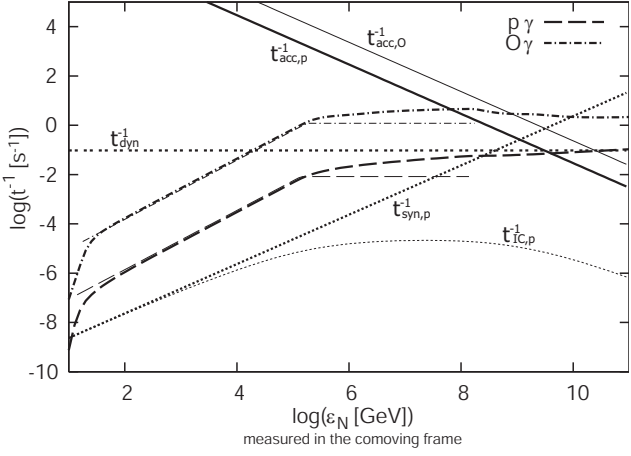}
\caption{\label{f1} The acceleration time scale and various cooling
time scales of proton and oxygen in the internal shock model for HL
GRBs. Energy and time scales are measured in the comoving frame of the
outflow. Used parameters are $L_{b}={10}^{51.5} \, \rm{erg} \,
{s}^{-1}$,  $\varepsilon_{\rm{ob}}^b ={10}^{2.5}$ keV, 
$\Gamma={10}^{2.5}$, $r={10}^{14}$ cm and $\xi_{B} (\approx 
\epsilon_B/\epsilon_e)=1$. Thick lines show numerical results on the
photomeson and/or photodisintegration time scales. Thin lines show
analytic results obtained by the resonance approximation. In the high
energies, the effect of the non-resonant cross section becomes important.
Note that this parameter set implies that a
significant fraction of the energy carried by protons goes into neutrinos.  
}
\end{figure}
\begin{figure}[tb]
\includegraphics[width=0.97\linewidth]{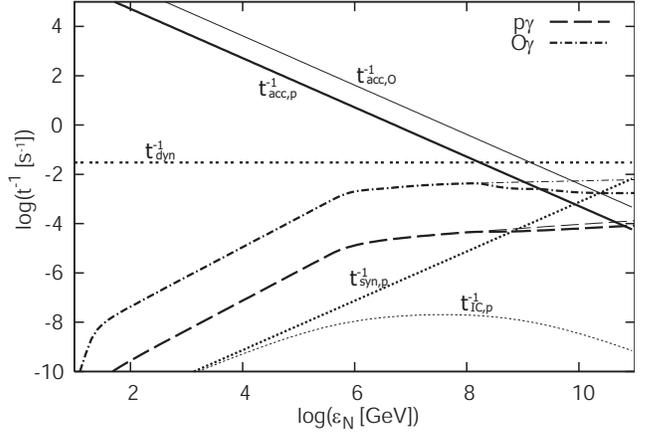}
\caption{\label{f2} The same as Fig. 1. But used parameters are 
$L_{b}={10}^{51} \, \rm{erg} \, {s}^{-1}$, $\varepsilon_{\rm{ob}}^b
={10}^{2.5}$ keV, $\Gamma={10}^{3}$, 
$r={10}^{15}$ cm and $\xi_{B} (\approx 
\epsilon_B/\epsilon_e)=1$. Thick lines show numerical results with 
the synchrotron self-absorption cutoff $1$ eV (in the comoving
frame). Thin lines show analytic results without the low-energy cutoff
for comparison. We can see that the effect of self-absorption
is not so important for the maximum energy due to the effect of the 
non-resonant cross section in the high energies. Note that this parameter set 
allows for survival of UHE nuclei.}
\end{figure}

The internal shock model is one of the most frequently discussed models in 
order to explain the prompt emission. Relativistic outflows make
internal collisions, which lead to internal dissipation via shocks. The formed
shocks will be mildly relativistic shocks, and charged particles will 
be accelerated at those collisionless shocks by some mechanism such as
the Fermi acceleration mechanism. HL GRBs have luminosity 
$L_{\gamma}\sim 10^{51-52} \, \rm{ergs}/\rm{s}$ and the observed
peak energy $\varepsilon_{\rm{ob}}^{b} \sim {10}^{2.5}$ keV. The
isotropic equivalent energy is $E_{\gamma}^{\mr{iso}} \sim {10}^{53}
\, {\rm{ergs}} \, L_{\gamma,52} [{\delta t}_{-1}/(1+z)] {N}_{2}$. 
Here $\delta t$ is the variability time which can vary in the very
wide range and $N $ is the number of collisions. A typical collision
radius will be expressed by commonly used relation, 
$r \approx 1.2 \times {10}^{14}{\Gamma}_{2.5}^2 {[\delta t/0.02 (1+z)
\, \mr{s}]}$ cm. Of course, this radius has to be smaller than the 
deceleration radius where the afterglow begins. As discussed in Refs. 
\cite{Mur1,Mur5,Asa1}, an internal collision radius is one of the most 
important quantities for the photomeson production. A collision radius $r \sim
{10}^{13-15}$ cm is frequently assumed. The magnetic field is given by $B = 7.3
\times {10}^{4} \, \mr{G} \epsilon _{B}^{1/2} {(\Gamma _{\mr{sh}}
(\Gamma _{\mr{sh}}-1)/2)}^{1/2} L _{\mr{M},52}^{1/2}
{\Gamma}_{2.5}^{-1} r_{14}^{-1}$. Here, $L_{\rm{M}}$ is the kinetic 
luminosity of outflows and $\Gamma_{\rm{sh}}$ is the relative Lorentz
factor between two subshells. The typical width of subshells in the comoving
frame $l$ is typically given by $l=r/\Gamma$.  

Next, let us evaluate maximum energies of cosmic rays by using several 
criteria. First, $t_{\rm{acc}}=t_{\rm{dyn}} \approx t_{\rm{ad}}$ leads to
\begin{eqnarray}
(1+z)E_{N,\rm{ad}}^{\rm{max}} &=& \frac{\Gamma ZeBl}{\eta} \nonumber \\
&\simeq& 6.9 \times {10}^{20} \, \mr{eV} \, Z \eta^{-1}
\epsilon _{B}^{1/2} \epsilon_e^{-1/2} \nonumber \\ &\times& {\left[ 
\frac{\Gamma_{\mr{sh}}(\Gamma_{\mr{sh}}-1)}{2} \right]}^{1/2} \!\!\!
L_{\gamma,51}^{1/2} {\Gamma}_{2.5}^{-1}. 
\end{eqnarray}
Note that the Hillas condition $r_L = l = r/\Gamma$ is already satisfied.
Second, $t_{\rm{acc}}=t_{\rm{syn}}$ leads to
\begin{eqnarray}
(1+z)E_{N,\rm{syn}}^{\rm{max}} &=& \sqrt{ \frac{6 \pi Ze}
{Z^4 \sigma _{T} B \eta}}\frac{\Gamma m_{N}^2
c^2}{m_e} \nonumber \\
&\simeq& 4.2 \times {10}^{20} \, \mr{eV} \, A^2 Z^{-3/2} \eta^{-1/2}
\epsilon _{B}^{-1/4} \epsilon_e^{1/4} \nonumber \\ &\times& {\left[ 
\frac{\Gamma_{\mr{sh}}(\Gamma_{\mr{sh}}-1)}{2} \right]}^{-1/4}
L_{\gamma,51}^{-1/4} {\Gamma}_{2.5}^{3/2} r_{14}^{1/2}. 
\end{eqnarray}
Therefore, we can expect that cosmic rays can be accelerated up to
ultra-high energies in the internal shock model of HL GRBs unless 
other cooling time scales such as $t_{p \gamma}$ are important.

The inverse Compton cooling time scale can be also calculated. 
For evaluation of $t_{\rm{IC}}$, we need to give a photon spectrum.
The photon spectrum for the prompt emission is often approximated 
by the broken power-law as 
\begin{equation}
\frac{dn}{d\varepsilon} = \frac{L_b
e^{-(\varepsilon/{\varepsilon}^{\rm{max}})}}{4 \pi r^2 \Gamma^2 c
{(\varepsilon^b)}^2} \left\{ \begin{array}{rl}
{(\varepsilon/\varepsilon^{b})}^{-\alpha} 
& \mbox{(for $\varepsilon ^{\mr{min}} \leq \varepsilon 
< \varepsilon ^b$)}\\
{(\varepsilon/\varepsilon^b)}^{-\beta} 
& \mbox{(for $\varepsilon ^b \leq \varepsilon \leq \varepsilon ^{\mr{max}}$)} 
\end{array} \right.
\end{equation}
where $L_b$ is the luminosity at the break energy measured by the 
observer in the local rest frame, $\varepsilon ^{\mr{min}}$ is the
minimum cutoff due to synchrotron self-absorption and 
$\varepsilon ^{\mr{max}}$ is the maximum cutoff due to pair-creation
in the comoving frame. In this paper, we set $\varepsilon
^{\mr{min}}=1$ eV and $\varepsilon^{\mr{max}}=10$ MeV. 
In this paper, we shall use $\alpha=1$ and $\beta=2.2$ as photon
indices. Although $t_{\rm{IC}}$ is calculated, we can usually ignore 
this cooling time scale due to the Klein-Nishina suppression. 
Hence, $E_{N,\rm{IC}}^{\rm{max}}$ is usually larger than 
$E_{N,\rm{syn}}^{\rm{max}}$.

The effect of the photomeson production process of protons is
investigated in detail in Ref. \cite{Mur1}. For details, 
see Refs. \cite{Mur1,Mur4} and references there in. At smaller
collision radii, $t_{p \gamma}$ can be more important than other
cooling time scales such as $t_{\rm{syn}}$. UHE protons are not
depleted only at sufficiently large radii. The cross section of 
photodisintegration is larger than that of photomeson production,
so that survival of UHE nuclei is more difficult than survival of UHE protons. 
We evaluate the maximum energy due to photomeson production and/or 
photodisintegration, $E_{p,p\gamma}^{\rm{max}}$ and/or 
$E_{N,N\gamma}^{\rm{max}}$, from $t_{\rm{acc}}=t_{p\gamma}$ and/or 
$t_{\rm{acc}}=t_{N\gamma}$.

Our numerical results on various time scales in the internal shock model
for HL GRBs are shown in Figs. 1 and 2.
For calculations, we give the total photon energy density 
$U_{\gamma}$ by $U_{\gamma} = \int d \varepsilon \, \varepsilon 
(\frac{dn}{d \varepsilon})$. The magnetic field is given by 
$U_B \equiv \xi_B U_{\gamma} \approx (\epsilon_B/\epsilon_e) U_{\gamma}$.
Fig. 1 shows the result for $r={10}^{14}$ cm and $\Gamma={10}^{2.5}$. 
Thick lines show numerical results while thin lines show curves when
we use resonance approximations. At sufficiently high energies, we can 
see that effects of the cross section in the non-resonance region 
become important.  
For this typical parameter set, we have the effective optical 
depth for the photomeson process $f_{p \gamma} \sim 0.3$. 
Hence, the efficient neutrino production occurs in this parameter
set, which can lead to detectable neutrino signals \cite{Wax2,Mur1}. 
However, we cannot expect survival of UHE
nuclei in such cases. Although they can be accelerated up to very high 
energies, UHE nuclei cannot survive in the sense that $f_{N \gamma} \gtrsim 1$.

Fig. 2 shows the result for $r={10}^{15}$ cm and $\Gamma={10}^{3}$.
In this parameter set, we expect that UHECR nuclei can survive. 
Protons and oxygens can be accelerated up to $\sim {10}^{20}$ eV and 
$\sim {10}^{21}$ eV, respectively.
This is just because the photon density becomes small enough at large
radii. Note that we have the effective optical depth for the photomeson
process $f_{p \gamma} \sim {10}^{-3}$, which is a
very small value. In this parameter set, although we can 
expect survival of UHE nuclei, the magnetic field is also expected to
be rather weak as long as we use the conventional value $\xi_B=1$. 
Here, we have $B \sim 5.8 \times {10}^{2}$ G, which seems to be
somewhat insufficient to explain the observed break energy by the
optically thin synchrotron model. (Then, we expect
$\varepsilon_{\rm{ob}}^b \sim 12 \, {\rm{keV}} \, \Gamma_{2.5} 
{(\Gamma_{\rm{sh}}-1)}^{2} \epsilon_{e}^2 g^2 f_{e}^{-2} B_{3}$, where
$g=g(p)=(p-2)/(p-1)$ and $p$ is the electron spectral index). 
But the detailed mechanism of the prompt emission has not been revealed yet
\cite{ioka07}, so that we do not consider this problem in this paper.
In Fig. 2, we also show 
curves for spectra with the minimum cutoff 
energy (thick lines) 
and that without the cutoff (thin
lines). Wang et al. \cite{Wan2} suggested that the effect of minimum cutoff
energy (due to self-absorption) can increase the maximum energy of
heavy nuclei. However, we see that the effect of self-absorption is not
so important even at the highest energies due to the effect of the
cross section in the non-resonance region, especially where the fragmentation
process is important.

\begin{figure}[tb]
\includegraphics[width=0.97\linewidth]{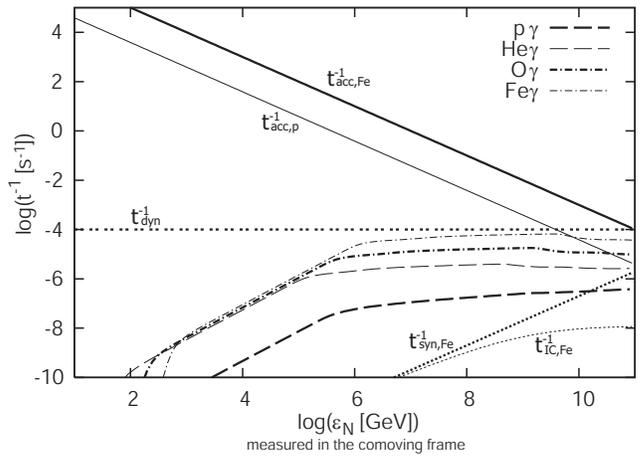}
\caption{\label{f3} The acceleration time scale and various cooling
time scales of proton, helium, oxygen and iron in the internal shock 
model for LL GRBs. Energy and time scales are measured in the comoving 
frame of the outflow. Used parameters are $L_{b}=1.5 \times {10}^{46} 
\, \rm{erg} \, {s}^{-1}$, $\varepsilon_{\rm{ob}}^b =5$ keV,
$\Gamma=10$, $r=6 \times {10}^{15}$ cm (corresponding to 
$T={10}^{3}$ s) and $\xi_{B} (\approx \epsilon_B/\epsilon_e)=1$. 
Note that this parameter set allows for survival of UHE nuclei.}
\end{figure}
\begin{figure}[tb]
\includegraphics[width=0.97\linewidth]{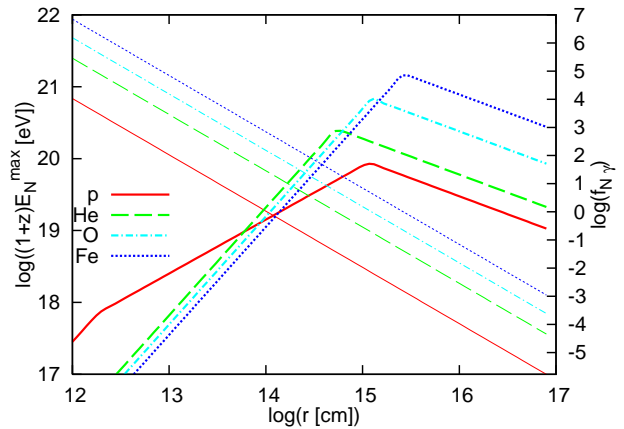}
\caption{\label{f4} The maximum energy of cosmic rays (thick lines)
and the optical depth for photomeson production or photodisintegration
(thin lines) as a function of the collision radius. Used parameters are 
$L_{b}=1.5 \times {10}^{46} \, \rm{erg} \, {s}^{-1}$, $\varepsilon_{\rm{ob}}^b
=5$ keV, $\Gamma=10$ and $\xi_{B} (\approx \epsilon_B/\epsilon_e)=1$ 
for LL GRBs.
The comoving shell width is set to $l=r/\Gamma$. Note that UHE nuclei 
can be produced in LL GRBs for $r \gtrsim {10}^{15}$ cm. There, UHE
nuclei can survive. At sufficiently large radii, $E_{N}^{\rm{max}}$
becomes $E_{N,\rm{ad}}^{\rm{max}}$.}
\end{figure}

Next, let us consider LL GRBs such as GRB 060218 and GRB 980425.  
For example, GRB 060218 has low luminosity $L_{\gamma} \sim 10^{46-47}$
ergs/s, which is much smaller than that of usual HL-GRBs \cite{Cam1}. 
The duration time is $T \sim 3000$ s, and the isotropic equivalent 
energy is $E_{\gamma}^{\mr{iso}} \sim {10}^{50}$ ergs. The observed
peak energy is $\varepsilon_{\rm{ob}}^b \sim 5$ keV, hence this event 
is classified as a x-ray flash.
Various interpretations for GRB 060218 exist. Several 
authors suggested the shock break-out model
\cite{Wax4,Wan3}. Other authors argued that this event can
be explained by the internal-external shock model 
\cite{Fan1,Ghi1,Tom1}. In this paper, we adopt the conventional internal shock
model following Toma et al. and we shall adopt the Lorentz
factor $\Gamma \sim (5-10)$, which is suggested in Refs. \cite{Ghi1,Tom1}.  

Fig. 3 shows one of our results for $r=6 \times {10}^{15}$ cm and
$\Gamma=10$, which correspond to $\delta t \sim 1000$ s when we use 
$r \approx 2 \Gamma^2 c \delta t$. Results of other parameter sets are
presented in Appendix A. In this parameter set demonstrated in Fig. 3,
we can expect that not only UHECRs can be produced but also
accelerated UHE nuclei can survive. 
This is just because the photon density is small enough at large
radii. Note that we have the effective optical depth for the photomeson
process $f_{p \gamma} \sim 2 \times {10}^{-3}$, which
is a small value. In this parameter set, although we can 
expect survival of UHE nuclei, the magnetic field seems to be 
somewhat weak when we use the conventional value $\xi_B=1$. In Fig. 3, 
Fe nuclei can be accelerated up to $\sim {10}^{21}$ eV, while protons
only up to $\sim {10}^{19.5}$ eV. 

As noted before, one of the important parameters is a collision
radius. Hence, we show maximum energies of proton and heavy nuclei 
$E_{N}^{\rm{max}}$ as a function of a collision radius $r$ in Fig. 4. 
In the same figure, $f_{p \gamma}$ and $f_{N\gamma}$ are shown. 
We can see that the
maximum energy of proton is determined by the photomeson production
at sufficiently small radii, but usually determined by the synchrotron
cooling and adiabatic cooling processes at large radii
($r \gtrsim {10}^{12.5}$ cm).
The maximum energy of nucleon is usually determined by the
photodisintegration and adiabatic cooling. Survival of UHE nuclei is
possible only at $r \gtrsim {10}^{15}$ cm in this parameter set.

\subsection{Reverse Shock Model}
The expanding fireball strikes the surrounding medium and
will form a reverse shock and forward shock. 
The shocked ambient and ejecta materials are in 
pressure balance and are separated by a contact discontinuity.
In the original standard model, the reverse shock is thought to be
short-lived, which exists during the initial deceleration of the
fireball. During this phase, optical/infrared flashes were expected.
Indeed, some optical flashes can be interpreted as 
the reverse shock emission. However, recent observations have reported 
the tentative lack of bright optical/infrared flashes \cite{Rom1}. 
Although it is one of the open problems in the \textit{Swift} era, we
do not manage this problem in this paper. Recently, the modern version of the
reverse shock model was developed to explain the shallow decay emission
in the early afterglow phase \cite{Gen1,Uhm1}. In such models, 
the plateau shape can be achieved by requiring the appropriate 
distribution of Lorentz factors of the ejecta and the 
suppression of the forward shock emission. For the 
reverse shock emission to emerge in the x-ray band, we can consider a number 
fraction of the shocked electrons that are injected into the
acceleration process $f_e^r$ to be smaller
than the unity. In this paper, we shall consider cosmic-ray
acceleration in the early afterglow phase under the reverse shock
model developed for the shallow decay emission, and adopt the small 
$f_e^r$ following Refs. \cite{Mur4,Gen1}.

The cosmic-ray production in the reverse shock model was discussed in
Refs. \cite{Vie1,Wax3,Dai1}. The more detailed study can be found in Ref. 
\cite{Mur4}. As demonstrated in these references, 
cosmic-ray acceleration up to ultra-high energies at a reverse shock
of HL GRBs is possible. In this paper, let us demonstrate that 
UHE cosmic-ray production at a reverse shock of LL GRBs is also possible.
First, $t_{\rm{acc}}=t_{\rm{dyn}} \approx t_{\rm{ad}}$ leads to 
\begin{eqnarray}
(1+z)E_{N,\rm{ad}}^{\mr{max}} = \frac{Z e B_{\times}^r r_{\times}}
{\eta} &\simeq& 4.7 \times {10}^{20} \, \mr{eV} \, Z \eta^{-1}
\nonumber \\
&\times& \epsilon_{B}^{1/2} E_{\mr{ej},50}^{3/8} n_2^{1/8} T_3^{1/8},
\end{eqnarray}
where $T$ is the duration time measured by the observer in the local
rest frame. Here, we have assumed the thick ejecta case, where 
the total ejecta thickness $\sim cT$ is larger than
the shocked region at the crossing time $\sim r_{\times}/2\Gamma_0^2$.
Here $\Gamma_0$ is the initial Lorentz factor.
In the thin ejecta case, the corresponding expression can be derived easily.
Second, $t_{\rm{acc}}=t_{\rm{syn}}$ leads to
\begin{eqnarray}
(1+z)E_{N,\rm{syn}}^{\mr{max}} &=& \sqrt{\frac{6 \pi Ze}
{Z^4 \sigma _{T} B_{\times}^r \eta}}\frac{\Gamma m_{N}^2
c^2}{m_e} \nonumber \\
&\simeq& 1.3 \times {10}^{21} \, \mr{eV} \, A^2 Z^{-3/2} \eta^{-1/2}
\nonumber \\ &\times& 
\epsilon_{B}^{-1/4} E_{\mr{ej},50}^{-3/16} n_2^{-5/16} T_3^{-1/16}.
\end{eqnarray}
Therefore, we can expect that UHECR production at a reverse shock is
possible not only for HL GRBs but also LL GRBs. Note that even protons 
could be accelerated above ${10}^{20}$ eV unless $\epsilon_B$ is too small.

Photon spectra in the reverse shock model can be calculated by 
exploiting the reverse-forward shock model. For example, in the 
slow cooling regime ($\varepsilon^{m}< \varepsilon^{sa}< \varepsilon^{c}$),
we obtain
\begin{equation}
\frac{dn}{d\varepsilon} = n_{\varepsilon,{\rm{max}}} 
e^{-(\varepsilon/{\varepsilon}^{\rm{max}})}
\left\{ \begin{array}{ll}
{(\varepsilon^{sa}/\varepsilon^{m})}^{-\frac{p+1}{2}}
{(\varepsilon^{m}/\varepsilon ^{sa})}^{\frac{3}{2}} 
{(\varepsilon/\varepsilon ^{m})}^{1} 
\\
{(\varepsilon^{sa}/\varepsilon^{m})}^{-\frac{p+1}{2}}
{(\varepsilon/\varepsilon ^{sa})}^{\frac{3}{2}} 
\\
{(\varepsilon/\varepsilon^{m})}^{-\frac{p+1}{2}} 
\\
{(\varepsilon^{c}/\varepsilon^{m})}^{-\frac{p+1}{2}}
{(\varepsilon/\varepsilon^c)}^{-\frac{p+2}{2}} 
\end{array} \right. \label{RSspectrum}
\end{equation}
where \cite{Mur4,Typ1}
\begin{equation}
n_{\varepsilon, \mr{max}}= \frac{L_{\varepsilon,\mr{max}}}{4 \pi
 r_{\times}^2 c {\varepsilon}^{n}}.
\end{equation}
Here $\varepsilon^n \equiv {\rm{min}}[\varepsilon^m,\varepsilon^c]$, and
$\varepsilon^m$, $\varepsilon^c$ and $\varepsilon^{sa}$ are 
the injection, cooling and self-absorption energy 
in the comoving frame, respectively. 
$L_{\varepsilon,\rm{max}}$ is the comoving specific luminosity per
unit energy at the injection or cooling energy.  
From Eq. (\ref{RSspectrum}), we can calculate $t_{\rm{IC}}$ and 
$t_{N\gamma}$. 

We show the numerical result of the reverse shock model for LL GRBs
in Fig. 5. The used parameters are $E_{\rm{ej}} = 2 \times {10}^{51}$
ergs, $n={10}^{2} \, \rm{cm}^{-3}$, $\epsilon_e^r=\epsilon_B^r=0.1$, 
$f_{e}^{r}=0.025$, $T=3000$ s and the initial Lorentz factor
$\Gamma_0=5$. As seen in Fig. 5, the UHECR production is
possible, where protons and Fe nuclei can be accelerated up to 
$\sim {10}^{20.2}$ eV. But we cannot expect survival of UHE nuclei in
this parameter set due to the copious photon field. For UHE nuclei
accelerated above ${10}^{20}$ eV to survive, for example, much smaller 
$\epsilon_e^r$ is needed. If we adopt such parameters, we can expect
both UHE protons and nuclei in this model.
\begin{figure}[tb]
\includegraphics[width=0.97\linewidth]{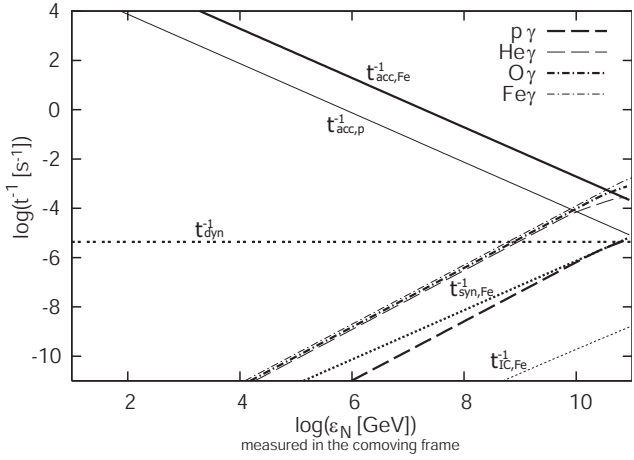}
\caption{\label{f5} The acceleration time scale and various cooling
time scales of proton, helium, oxygen and iron in the reverse shock 
model for LL GRBs. Energy and time scales are measured in the comoving 
frame of the outflow. Used parameters are $E_{\rm{ej}}=2 \times 
{10}^{51} \, \rm{ergs}$, $n={10}^{2} \, {\rm{cm}}^{-3}$ (ISM), 
$\Gamma_0=5$, $T= 3 \times {10}^{3}$ s, $\epsilon_{B}^r=0.1$, 
$\epsilon_e^r=0.1$, $f_e^r=0.025$ and $p=2.4$. Note that this parameter 
set does not allow for survival of UHE nuclei.}
\end{figure}

\subsection{Forward Shock Model}
\begin{figure}[tb]
\includegraphics[width=0.97\linewidth]{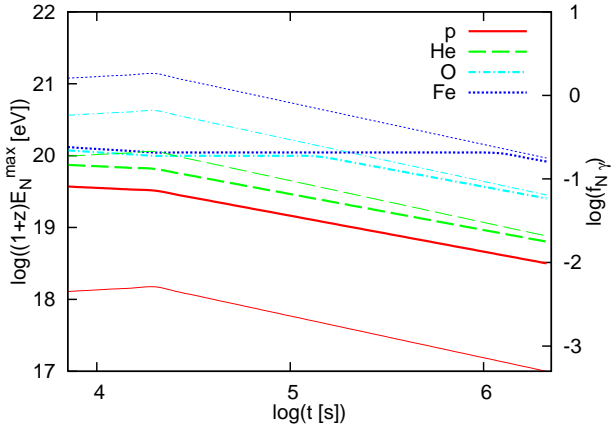}
\caption{\label{f6} The maximum energy of cosmic rays (thick lines)
and the optical depth for photomeson production or photodisintegration
at ${10}^{20}$ eV (thin lines) as a function of the time after the burst. Used 
parameters are $E_{\rm{ej}}=2 \times {10}^{51} \, \rm{ergs}$, 
$n={10}^{2} \, {\rm{cm}}^{-3}$ (ISM), $\epsilon_{B}=0.01$, 
$\epsilon_e=0.01$, $f_e=0.1$ and $p=2$. The jet-break time is set to 
$t_j= 2 \times {10}^{4}$ s. Note that the deceleration time is 
$t_{\rm{dec}} \simeq 8000$ s. UHE nuclei can be produced in a forward 
shock of LL GRBs. Survival of UHE nuclei becomes the most difficult at
the jet-break time. At sufficiently late time, $E_{N}^{\rm{max}}$ becomes 
$E_{N,\rm{ad}}^{\rm{max}}$. }
\end{figure}
\begin{figure}[tb]
\includegraphics[width=0.97\linewidth]{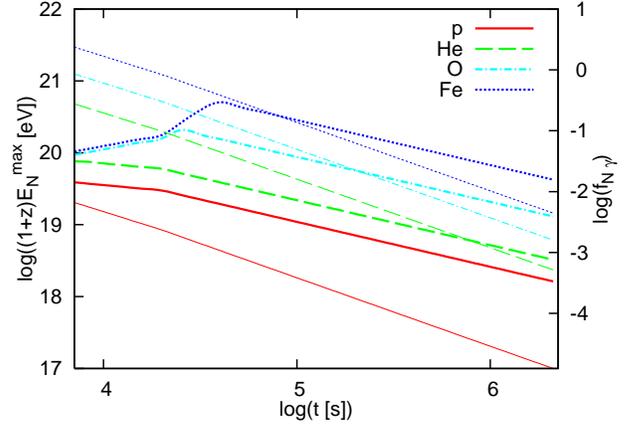}
\caption{\label{f7} The same as Fig. 6. But the wind-like circumburst
medium is assumed ($A_*=0.77$). UHE nuclei can be produced in a forward 
shock of LL GRBs. Survival of UHE nuclei becomes easier with time.  
At sufficiently late time, 
$E_{N}^{\rm{max}}$ becomes $E_{N,\rm{ad}}^{\rm{max}}$.}
\end{figure}
The expanding fireball strikes the surrounding medium and will 
form a forward shock. The self-similar behavior after the deceleration 
time $t_{\rm{dec}}$ is described by the Blandford-Mckee solution. The standard
external shock model based on this solution can reproduce observations
at the late time. Although the original forward shock model actually 
fails to explain the shallow decay emission, we consider the original
forward shock model for simplicity. By exploiting the theory, we
immediately have $\Gamma \propto t^{-\frac{3}{8}}$, 
$r \propto t^{\frac{1}{4}}$ and $B \propto  t^{-\frac{3}{8}}$. 
By using them, we can derive the time-dependence of various quantities.
Typical parameters for HL GRBs are $E_{\rm{ej}} \sim {10}^{52-53}$ ergs, 
$n=1 \, {\rm{cm}}^{-3}$ (ISM), $\epsilon_B \sim 0.01$, 
$\epsilon_e \sim 0.1$ and $p \sim 2$.
In obtaining above parameters, all the electrons are often assumed to be
accelerated. However, this may not be true. As Eichler \& Waxman
\cite{Eic1} discussed, only a fraction $f_e$ of electrons may be
accelerated, that is, $f_e$ can be smaller than the unity. 
When we consider the small $f_e$, the kinetic energy of ejecta can be
larger by $1/f_e$. Such small values of $f_e$ will be tested by 
observations of polarization in the future \cite{Tom2}.

Cosmic-ray acceleration and neutrino production in the forward shock 
model for HL GRBs were discussed in detail by Dermer \cite{Der2,Der4}.
Hereafter, let us demonstrate UHECR production at a forward shock for
parameters of LL GRBs. In this paper, we shall adopt 
$E_{\rm{ej}}=2 \times {10}^{51} \, \rm{ergs}$, 
$n={10}^{2} \, {\rm{cm}}^{-3}$ (ISM), $\epsilon_{B}=0.01$, 
$\epsilon_e=0.01$, $f_e=0.1$ and $p=2$. 
This parameter set is consistent with parameters obtained by Toma et
al. \cite{Tom1} except that we use $f_e=0.1$.

Next, let us estimate maximum energies of cosmic rays.
First, the maximum energy determined from the condition 
$t_{\rm{acc}}=t_{\rm{dyn}} \approx t_{\rm{ad}}=$ is \cite{Gal1,Ach1,Mur4}
\begin{eqnarray}
(1+z)E_{N,\rm{ad}}^{\mr{max}} &=& \frac{Z e  \Gamma B^{\mr{ISM}}
r}{\eta} \nonumber \\ &\simeq& 7.1 \times {10}^{14} \, \mr{eV} \,
Z \eta^{-1} \nonumber \\ 
&\times& B_{-6}^{\mr{ISM}} E_{\mr{ej},51}^{3/8} n_2^{-3/8}{t_4}^{-1/8}. 
\end{eqnarray}
Here $(1+z)t$ is the time after the prompt emission measured by the observer.
When we use the conventional value
of the ISM magnetic field, we cannot expect UHECR production at a
forward shock of GRBs.
However, the upstream magnetic field might be amplified significantly 
via some process, e.g., the non-resonant cosmic-ray streaming instability
\cite{Bel1,Der2} (but see also Ref. \cite{Mil2}). Instead, if the 
second-order Fermi acceleration works well \cite{Der5}, we could 
use the magnetic field in the downstream, which is likely to be
amplified, as indicated by observations.
If the upstream and/or downstream magnetic fields are amplified up 
to near the equipartition value, $t_{\rm{acc}}=t_{\rm{dyn}} \approx 
t_{\rm{ad}}$ leads to
\begin{eqnarray}
(1+z)E_{N,\rm{ad}}^{\mr{max}} = Z e B^{f} r \eta^{-1} 
&\simeq& 2.7 \times {10}^{20} \, \mr{eV} \, Z \eta^{-1} \nonumber \\
&\times& \epsilon_{B}^{1/2} E_{\mr{ej},51}^{3/8} n_2^{1/8} 
t_4^{-1/8}\!\!\!\!. \label{FSEad}
\end{eqnarray} 
The maximum energy limited by synchrotron cooling is
\begin{eqnarray}
(1+z)E_{N,\rm{syn}}^{\mr{max}} &=& \sqrt{\frac{6 \pi Ze}
{Z^4 \sigma _{T} B^f \eta}}\frac{\Gamma m_{N}^2
c^2}{m_e} \nonumber \\
&\simeq& 2.5 \times {10}^{20} \, \mr{eV} \, A^2 Z^{-3/2} \eta^{-1/2}
\nonumber \\ &\times& 
\epsilon_{B}^{-1/4} E_{\mr{ej},51}^{1/16} n_2^{-5/16} t_4^{-3/16}. 
\label{FSEsyn}
\end{eqnarray}
From Eqs. (\ref{FSEad}) and (\ref{FSEsyn}), we can expect that UHECR 
production is also possible at a forward shock of LL GRBs.

A photon spectrum in the forward shock model can be also calculated
from the theory. For example, in the slow cooling case 
($\varepsilon^{sa}< \varepsilon^{m}< \varepsilon^{c}$), we have
\begin{equation}
\frac{dn}{d\varepsilon} = \frac{L_{\varepsilon,\mr{max}} 
e^{-(\varepsilon/{\varepsilon}^{\rm{max}}})}{4 \pi
 r^2 c {\varepsilon}^{n}}
\left\{ \begin{array}{ll}
{(\varepsilon^{sa}/\varepsilon^{m})}^{-\frac{2}{3}}
{(\varepsilon/\varepsilon ^{sa})}^{1} 
\\
{(\varepsilon/\varepsilon ^{m})}^{-\frac{2}{3}} 
\\
{(\varepsilon/\varepsilon^{m})}^{-\frac{p+1}{2}} 
\\
{(\varepsilon^{c}/\varepsilon^{m})}^{-\frac{p+1}{2}}
{(\varepsilon/\varepsilon^c)}^{-\frac{p+2}{2}} 
\end{array} \right. \label{FSspectrum}
\end{equation}
Here, $\varepsilon_{\rm{ob}}^m = \Gamma \varepsilon^m \simeq 0.17 
\, {\rm{eV}} \, g_{-1}^2  f_{e,-1}^{-2} \epsilon_{B,-2}^{1/2} \epsilon_{e,-1}^2
E_{\rm{ej},51}^{1/2} t_{4}^{-3/2}$, $\varepsilon_{\rm{ob}}^c =
\Gamma \varepsilon^c \simeq 1.0 \, {\rm{eV}} \, \epsilon_{B,-2}^{-3/2} 
E_{\rm{ej},51}^{-1/2} n_2^{-1} t_{4}^{-1/2}$ and 
$L_{\varepsilon_{\rm{ob}},\rm{max}} = \Gamma L_{\varepsilon, \rm{max}}
\simeq 7.8 \times {10}^{56} \, {\rm{s}}^{-1} \, f_{e,-1} \epsilon_{B,-2}^{1/2} 
E_{\rm{ej},51} n_2^{1/2} (\phi_p/0.6)$, where $\phi_p$ is an
order-of-unity factor calculated by Wijers \& Galama \cite{Wij1}.

The photodisintegration and/or photomeson time scales are 
evaluated from Eq. (\ref{tpg}). Here, let us estimate 
these time scales analytically in the slow cooling case by applying Eq. 
(\ref{tNg2}). After replacing $\varepsilon^b$ with 
$\varepsilon^c$, resonance approximations give
\begin{eqnarray}
t_{N \gamma}^{-1} &\simeq&  g_{-1}(\phi_p/0.6) \epsilon_{e,-2} 
\epsilon_{B,-2}^{3/2} E_{\rm{ej},51}^{7/8} n_2^{13/8} t_{4}^{-5/8}
{(E_N/E_N^b)}^{1/2}\nonumber \\ &\times&
\left\{ \begin{array}{ll} 
{1.1 \times {10}^{-8} \, {\rm{s}}^{-1}} & \mbox{(for $\Delta$-resonance)}\\
{5.7 \times {10}^{-8} A^{1.21} \, {\rm{s}}^{-1}} & \mbox{(for GDR)}
\end{array} \right.
\end{eqnarray}
The above expressions are obtained for $E_{N}>E_N^b = 0.5 
\bar{\varepsilon}_{\rm{res}} m_N c^2 \Gamma^2 / \varepsilon_{\rm{ob}}^c$. 
Hence, we have $f_{N \gamma} \propto t^{\frac{1}{8}}$ and 
$E_{N, N\gamma}^{\rm{max}} \propto t^{-\frac{1}{6}}$ for 
$E_N > E_N^b$. Of course, the time-dependence differs for 
$E_N < E_N^b$. In the ISM case, we can see that $f_{N \gamma}$
increases with time. Note that the temporal index of $f_{N \gamma}$ is
different from that used in Wang et al. \cite{Wan2}, because we
consider $E_{N} \sim {10}^{20}$ eV, where $E_{N} > E_N^b$ is typically
expected rather than $E_{N} < E_N^b$ considered by them.  

In the wind-medium case, we have $E_{N,\rm{ad}}^{\rm{max}} \propto 
t^{-\frac{1}{4}}$ and $E_{N,\rm{syn}}^{\rm{max}} \propto
t^{\frac{1}{8}}$. Furthermore, we can derive $E_{N, N\gamma}^{\rm{max}} 
\propto t^{\frac{1}{2}}$ and $f_{N \gamma} \propto t^{-1}$ for $E_N >
E_N^b$. In the wind-medium case, we can see that $f_{N \gamma}$ 
decreases with time. It is because the ambient density
drops as $r^{-2}$, which leads to rapid drop of the photon density with time.

In the light curve of afterglows of HL GRBs, we can often find the jet
break around $t_{j} \sim {10}^{5}$ s.  
After the jet break, we have $r \propto t^0$, $\Gamma \propto
t^{-1/2}$ and $B \propto t^{-1/2}$. 
Therefore, for example, we obtain $E_{N,\rm{ad}}^{\rm{max}} \propto 
t^{-\frac{1}{2}}$, $E_{N,\rm{syn}}^{\rm{max}} \propto
t^{-\frac{1}{4}}$, $E_{N, N\gamma}^{\rm{max}} \propto t^{0}$ and 
$f_{N \gamma} \propto t^{-\frac{1}{2}}$ for $E_N > E_N^b$ in the ISM case.
After the jet break, the photon density will become smaller and smaller. 
Hence, we expect photodisintegration and photomeson production
become less important. 
Similarly, we can obtain the time-dependence of various quantities for 
the wind-medium case. 
In both cases, the effect of the jet break makes survival of UHE
nuclei easier. It has not been pointed out in previous studies.   

Results in the forward shock model for LL GRBs are shown in Figs. 6
and 7. In Fig. 6, we show maximum energies of cosmic rays
$E_{N}^{\rm{max}}$ and the optical depth for photodisintegration
$f_{N\gamma}$ at $E_{N}={10}^{20}$ eV as a function of time $t$. 
From this figure, we can expect UHE nuclei can be produced at a 
forward shock of LL GRBs. For heavy nuclei (O and Fe), the
photodisintegration determines the maximum energy at the
earlier time, but the adiabatic cooling becomes more important at 
the later time. For light nuclei, the adiabatic cooling is almost 
always the most important. Note that the jet-break effect can be
important for survival of UHE nuclei, as expected. In fact, after 
the jet break, $f_{{\rm{Fe}}\gamma}$ can decrease with time and become 
smaller than the unity.

In Fig. 7 we show the results for the wind-medium case, In this case,
we also expect acceleration and survival of UHE nuclei. 
Since the photon density decreases with time, the maximum
energy of heavy nuclei, which is determined by
$E_{N,N\gamma}^{\rm{max}}$, can increase until the adiabatic cooling
becomes more important. As a result, UHE nuclei could be accelerated up
to $\gtrsim {10}^{20}$ eV in this parameter set.

\section{Implications For Neutrino Astronomy}
Sufficiently high-energy cosmic rays cannot 
avoid the photomeson production process, where the pion production
threshold energy is $145$ MeV. Generated pions, kaons and other mesons 
can produce high-energy neutrinos. Such high-energy neutrino signals, 
if detected, are very important as a direct probe of cosmic-ray acceleration. 
High-energy neutrino emission from accelerated protons at shocks was 
predicted by Waxman \& Bahcall \cite{Wax2} in the internal shock model. 
Predictions were also done in the reverse-forward shock model
\cite{Wax3,Dai1,Der2}. Further predictions in the \textit{Swift}-era are found 
in Refs. \cite{Mur2,Mur3,Mur4}.  

The important quantity for the neutrino flux originating from protons 
is the effective optical depth for the photomeson production process 
$f_{p \gamma}$, which also represents an
energy fraction of protons carried by mesons, as long as $f_{p \gamma}<1$.   
When $f_{p \gamma} \gtrsim 1$, accelerated protons are depleted by
photomeson production, which also means the efficient photomeson
production. Such conditions are satisfied when internal shocks making 
the prompt emission occur at sufficiently small radii \cite{Mur1}. 
In addition, $f_{p \gamma} \gtrsim 1$ is also possible in the case of
flares and late prompt emissions \cite{Mur3,Mur4}. 

On the other hand, when $f_{p \gamma} \lesssim 1$,
high-energy protons can survive without complete depletion. Such
conditions are satisfied when internal shocks making 
prompt emission occur at sufficiently large radii. In the
reverse-forward shock model (with the ISM environment), 
$f_{p \gamma} \lesssim 1$ is usually expected except in the highest-energies. 
However, we still expect that a significant fraction of the nonthermal proton
energy is released as neutrinos unless $f_{p \gamma}$ is too small
\cite{Wax3,Der2,Mur4}.  

In most previous works, all the cosmic rays are assumed to be
protons. However, heavier nuclei may be entrained, which can also be 
accelerated up to ultra-high energies as shown in Sec. \Roman{ni}. 
We can relate $f_{p \gamma}$ with $f_{N \gamma}$ as 
\begin{eqnarray}
\frac{f_{p \gamma}(E_N)}{(f_{N \gamma}(E_N)/A)} \simeq 0.2 A^{-0.21} 
\left\{ \begin{array}{ll} 
{E_N}^{\beta-\alpha} {(E_p^b)}^{1-\beta} {(E_N^b)}^{\alpha-1}\\
{(E_N^b/E_p^{b})}^{\alpha-1} 
\end{array} \right. \label{pgandNg} 
\end{eqnarray}
Here, the parameter regions for the upper and lower columns 
are $(E_N^b \leq) \, E_N < E_p^b$ and $E_N \geq E_p^b$, respectively.
From Eq. (\ref{pgandNg}), we can see that $f_{p \gamma}$ is small at 
radii where UHE heavy nuclei can survive, $f_{N \gamma} < 1$.  

We can calculate neutrino spectra by using the same method as in
Ref. \cite{Mur1}. When UHE heavy nuclei such as iron can survive, 
we easily expect that neutrino emission is not efficient from Eq. 
(\ref{pgandNg}). For calculations, we assume that cosmic rays are all
protons in cases where UHE irons cannot survive, while proton 
75 \% and iron 25 \% in cases where UHE irons can survive.
Although this cosmic-ray composition is just ad hoc, our results on
neutrino spectra in GRBs do not depend on the composition so much,
especially when UHE heavy nuclei survive, $f_{N \gamma} < 1$.      
The reason is as follows. The cross section of photomeson production for 
heavy nuclei is roughly expressed as $\sigma_{\rm{meson}} \simeq A 
\sigma_{\Delta}$ (rather than $A^{2/3} \sigma_{\Delta}$ which is given
in Ref. \cite{Wan2}), while the inelasticity 
around the $\Delta$-resonance is around $0.2/A$. Hence, the energy
loss time scales due to the photomeson production of heavy nuclei is 
written as $t_{\rm{meson}} \approx t_{p \gamma}$. Hence, protons and irons
generate similar neutrino fluence levels in cases of GRBs, as long as both
are accelerated up to sufficiently high energies \cite{Com2,Anc1}.    

As an example, let us show neutrino spectra expected in the 
prompt emission phase, which is the most frequently discussed 
since the prediction by Waxman \& Bahcall \cite{Wax2}. 
We can evaluate $f_{p\gamma}$ and $f_{\rm{meson}}$ analytically
by using the $\Delta$-resonance approximation \cite{Wax2,Mur3}. For HL
GRBs, we obtain
\begin{equation}
f_{p\gamma} \approx f_{\rm{meson}} 
\simeq 0.3 \frac{L_{b,51.5}}{r_{14} {\Gamma_{2.5}}^2
\varepsilon_{\rm{ob},316 \, \mr{keV}}^{b}} \left\{ \begin{array}{rl} 
{(E_p/E_p^b)}^{\beta-1}\\
{(E_{p}/E_{p}^{b})}^{\alpha-1}
\end{array} \right. \label{pgHL}
\end{equation}
Here, the parameter regions for the upper and lower columns 
are $E_p < E_p^b$ and $E_p \geq E_p^b$, respectively, and we have 
multiplied a factor of $\sim 2.5$ due to the effect of
the multi-pion production which is important for $\alpha \sim 1$
spectra \cite{Mur3}. 
For LL GRBs, we can apply the above expression \cite{Mur2} and have
\begin{equation}
f_{p \gamma} \approx f_{\rm{meson}} \simeq 1.4 \times {10}^{-3} \frac{
L_{b,46.2}}{r_{15.8} {\Gamma_1}^2
{\varepsilon}_{\rm{ob}, 5 \, \mr{keV}}^{b}} \left\{ \begin{array}{rl} 
{(E_p/E_p^b)}^{\beta-1}\\
{(E_{p}/E_{p}^{b})}^{\alpha-1} 
\end{array} \right. \label{mesonLL}
\end{equation}
Here, the parameter regions for the upper and lower columns 
are $E_p < E_p^b$ and $E_p \geq E_p^b$, respectively.
Our results are shown in Fig. 8. In fact, the above analytic
estimations agree with numerical results. For example, let us consider
parameter sets demonstrated in Fig. 1 for HL GRBs and Fig. 3 for LL
GRBs. For the former set with the source redshift $z=0.1$ 
($E_{\gamma}^{\rm{iso}}={10}^{53}$ ergs and $\xi_{\rm{acc}}=20$), we have 
$E_{\nu}^2 \phi_{\nu} \sim (1/4) f _{p \gamma} E_p^2 (d
N_p^{\rm{iso}}/d E_p) / (4 \pi D^2) \sim 3 \times {10}^{-4} \,
\rm{erg} {cm}^{-2}$, which agrees with the thick solid line shown in
Fig. 8. For the latter set with the source redshift $z=0.005$ 
($E_{\gamma}^{\rm{iso}}={10}^{50}$ ergs and $\xi_{\rm{acc}}=10$), we have 
$E_{\nu}^2 \phi_{\nu} \sim (1/4) f _{p \gamma} E_p^2 (d
N_p^{\rm{iso}}/d E_p) / (4 \pi D^2) \sim 7 \times {10}^{-7} \, \rm{erg}
{cm}^{-2}$, which also agrees with the thin dashed line shown in
Fig. 8. Note that such low redshift bursts (at $\sim 20$ Mpc) have not
been observed yet (e.g., $\sim 40$ Mpc for GRB 980425). But we may see
such bursts if LL GRBs occur in e.g., Virgo cluster. The expected muon 
event rates by IceCube are also shown in the figure caption of Fig. 8.       
As stressed in the previous paragraph, survival of UHE heavy
nuclei means that neutrino emission is inefficient, so that it would
be difficult to expect detection of neutrino signals by near-future
neutrino telescopes such as IceCube.  
 
Since it is difficult to see neutrino signals from one GRB event, we may need
to see many neutrino events as the cumulative neutrino background. 
As we can see from Eqs. (\ref{nubkgHL}) and (\ref{nubkgLL}), the
cumulative neutrino flux can be estimated from 
${\rm{min}}[1,f_{p \gamma}]$ and a given cosmology (see Appendix C). 
We typically expect ${\rm{min}}[1,f_{p \gamma}] \sim (0.01-1)$, for
example, in the internal shock model for HL GRBs with 
$\Gamma \lesssim {10}^{2.5}$ and
$r \lesssim {10}^{15.5}$ cm. Smaller values are possible only at
larger radii and/or for larger Lorentz factors. Survival of UHE
heavy nuclei such as iron requires such relatively extreme parameter
sets, which leads to $f_{p \gamma} \sim {10}^{-3}$.
As a result, the expected cumulative neutrino
flux under the GRB-UHECR hypothesis is $E_{\nu}^{2} \Phi_{\nu} \sim {10}^{-8}
\, \rm{GeV} \rm{cm}^{-2} \rm{s}^{-1} \rm{sr}^{-1}$ for the parameter
set demonstrated in Fig. 1, while $E_{\nu}^{2} \Phi_{\nu} \sim 3 
\times {10}^{-11} \, \rm{GeV} \rm{cm}^{-2} \rm{s}^{-1} \rm{sr}^{-1}$
for the parameter set demonstrated in Fig. 2.
The corresponding muon event rates by IceCube are $N_{\mu} \sim 50$
events/yr and $N_{\mu} \sim 0.05$ events/yr, respectively. 
Since the neutrino flux from nuclei is very similar to that from protons
when accelerated heavy nuclei survive, we can use results obtained in
Murase \& Nagataki for mixed composition cases where UHE nuclei can
survive. The detailed numerical calculations on the cumulative neutrino
background are found in Refs. \cite{Mur1,Mur2,Mur3,Mur4}. In
Ref. \cite{Mur1}, neutrino spectra are shown for various collision
radii and it is useful to compare set A and set B in Figs. 15-17, for example.  
So far we have considered the internal shock model. For other models,
see Appendixes D and E.

\begin{figure}[tb]
\includegraphics[width=0.97\linewidth]{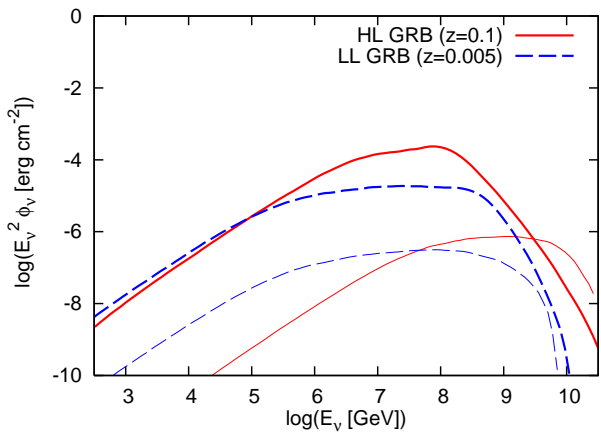}
\caption{\label{f8} Energy fluences of neutrinos from one nearby GRB
event. Solid lines and dashed lines show HL GRB with
$E_{\gamma}^{\rm{iso}}={10}^{53}$ ergs at $z=0.1$ and LL GRB with
$E_{\gamma}^{\rm{iso}}={10}^{50}$ ergs at $z=0.005$, respectively. 
A thick solid line shows the HL-GRB neutrino spectrum for $r={10}^{14}$
cm and $\Gamma = {10}^{2.5}$ where heavy nuclei cannot survive, while
a thin solid line shows the HL-GRB neutrino spectrum for $r={10}^{15}$
cm and $\Gamma = {10}^{3}$ where heavy nuclei can survive (see Figs. 1
and 2). A thick dashed line shows the LL-GRB neutrino spectrum for
$r=9 \times {10}^{14}$ cm and $\Gamma = {10}$ where heavy nuclei
cannot survive, while a thin dashed line shows the LL-GRB neutrino 
spectrum for $r=6 \times {10}^{15}$ cm and $\Gamma = {10}$ where heavy 
nuclei can survive (see Figs. 3 and 10). 
The cosmic-ray composition with proton 100 \% is assumed for
thick lines, while proton 75 \% and iron 25 \% for thin lines. 
The nonthermal baryon loading factors
$\xi_{\rm{acc}} \equiv U_{\rm{CR}}/U_{\gamma}$ are set to 20 for HL GRBs and
10 for LL GRBs, respectively (see Appendix B). We also use
$\xi_B\equiv U_B/U_{\gamma}=1$.
Expected muon event rates by IceCube are $N_{\mu} \sim 1$ event for
the thick solid line, $N_{\mu} \sim 0.001$ events for
the thin solid line, $N_{\mu} \sim 0.2$ events for
the thick dashed line and $N_{\mu} \sim 0.002$ events for
the thin dashed line.}
\end{figure}

\section{Implications for Gamma-Ray Astronomy}
Not only neutrinos but also high-energy gamma rays originating from
cosmic rays (cosmic-ray synchrotron radiation), neutral pions, and
muons, electrons and positrons from charged pions 
will be produced. However, such high-energy gamma rays generally
suffer from the internal attenuation processes, especially in the
internal shock model, as discussed in many papers (see, e.g.,
\cite{Mur5} and references there in). The copious photon
field also plays an important role on the efficient photomeson
production, so that we cannot expect that GRBs are bright in $\sim$TeV 
gamma rays when bright in neutrinos (see Refs. \cite{Der3,Der6} and 
references there in). In other words, when $f_{p \gamma}$ becomes 
small enough, we can expect that the optical
depth for pair-creation $f_{\gamma \gamma}$ becomes smaller than the
unity (hence high-energy gamma rays from far sources will be attenuated by 
CMB/CIB photons rather than seed photons in the
source). As seen in the previous section, when UHE heavy nuclei can
survive, $f_{p \gamma}$ is
much smaller than the unity. Therefore, we may expect escape of very
high-energy gamma rays from the source when UHE heavy nuclei can
survive. Such high-energy gamma-ray signals are useful as
signatures of UHECR acceleration, so that they are important although 
the distinction between hadronic and leptonic components is not so
easy in general. Note that, when $f_{p \gamma}$ is large enough, one 
may not expect high-energy gamma rays due to internal attenuation. 
But proton-induced cascaded gamma-ray signals could be seen, e.g., when the 
baryon loading factor is large enough $\xi_{\rm{acc}} \gg 10$ 
\cite{Der6,Asa2}. Hence, observations of high-energy gamma rays are 
still important as a probe of the UHECR acceleration. See also \cite{Typ5}.
  
As an example, let us consider the internal shock model with $\alpha=1$ and 
$\beta=2.2$. The optical depth for pair-creation process (which is
usually relevant for escape of high-energy gamma rays) can be
approximately written as $f_{\gamma \gamma} (\varepsilon_{\rm{ob}}) 
\simeq 0.12 \sigma_T \tilde{\varepsilon} {(dn/d
{\varepsilon})}_{\tilde{\varepsilon}} l$ \cite{Sve1,Mur5}. Here 
$\tilde{\varepsilon}$ is defined as ${(m_e c^2)}^{2}/\varepsilon$. 
When $\alpha \sim 1$, the optical depth for pair-creation 
$f_{\gamma \gamma}$ roughly reaches the maximum value at 
$\sim \tilde{\varepsilon}_{\rm{ob}}^b = \Gamma^2 
{(m_e c^2)}^2/\varepsilon_{\rm{ob}}^b$. (Note that, in fact, 
$f_{\gamma \gamma} (\varepsilon_{\rm{ob}})$ increases
logarithmically above $\sim \varepsilon_{\rm{ob}}^b$ while decreases 
above $\tilde{\varepsilon}_{\rm{ob}}^{sa}$.)
On the other hand, as seen in the previous section, $f_{N\gamma}$ is
also determined by the photon field. Therefore, we can relate the two
quantities. Note that, in cases of $\alpha \sim 1$, the numerically 
calculated $f_{N\gamma}$ will be effectively larger than 
that evaluated analytically by using Eq. (\ref{tNg2}) by a factor of 
$\sim 3$. Taking into account this factor, $f_{\gamma \gamma}$ at 
$\varepsilon_{\rm{ob}}$ can be written as 
\begin{eqnarray}
\frac{f_{\gamma \gamma} ({\varepsilon}_{\rm{ob}})}
{ f_{N \gamma}(E_N)} \simeq 0.95 {\left( \frac{A}{56} \right)}^{-1.21} 
 \left\{ \begin{array}{ll} {(\varepsilon_{\rm{ob}}
/{\tilde{\varepsilon}}_{\rm{ob}}^b)}^{1.2}\\
{\,\,\,\,\,\,\,\,\,\, 1} 
\end{array} \right. \label{ggandNg}
\end{eqnarray}
Here, the parameter regions for the upper and lower columns 
are $\varepsilon_{\rm{ob}} < \tilde{\varepsilon}_{\rm{ob}}^b$ and 
$\varepsilon_{\rm{ob}} \geq \tilde{\varepsilon}_{\rm{ob}}^b$, respectively. 
From the above expressions, we can expect that the cosmic-ray source
where UHE irons can survive ($f_{\rm{Fe} \gamma} \lesssim 1$) is 
almost completely thin for pair-creation ($f_{\gamma \gamma} \lesssim 1$). 
This implies that very high-energy gamma rays are expected in the prompt
phase, if UHE irons can be accelerated at internal shocks and survive. 
In Fig. 9, we demonstrate gamma-ray spectra from one HL GRB at $z=0.1$ and LL
GRB at $z=0.005$. Gamma-ray spectra given by Eq. (6) and originating 
from cosmic-ray synchrotron radiation are shown. The former usually
attributes to synchrotron (or jitter) radiation from accelerated electrons. 
The latter component should exist
if cosmic rays can be accelerated to ultra high energies. Very
high-energy gamma rays can be also produced via synchrotron
self-inverse Compton scattering, which should contaminate the 
cosmic-ray synchrotron component. Whether such a synchrotron
self-inverse Compton component exists or not,
we can expect to detect high-energy gamma-rays by GLAST and/or
MAGIC in the future, as long as accelerated UHE
heavy nuclei survive and GRBs occur at $z \lesssim 1$ for HL GRBs and 
$z \lesssim 0.05$ for LL GRBs, from Fig. 9.  

\begin{figure}[tb]
\includegraphics[width=0.97\linewidth]{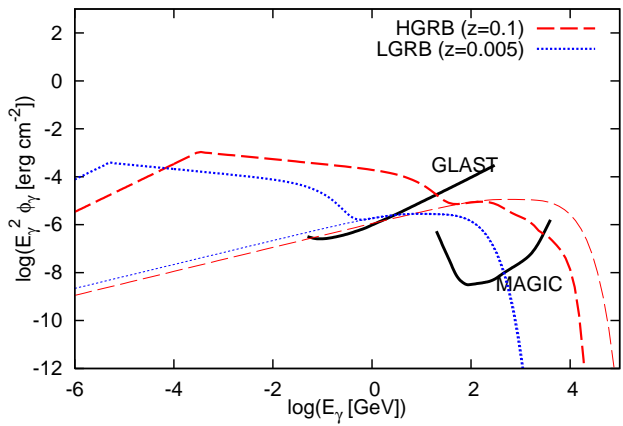}
\caption{\label{f9} Energy fluences of gamma rays from one nearby GRB
event. Dashed and dotted lines show HL GRB with
$E_{\gamma}^{\rm{iso}}={10}^{53}$ ergs at $z=0.1$ and LL GRB with
$E_{\gamma}^{\rm{iso}}={10}^{50}$ ergs at $z=0.005$, respectively. 
A thick dashed line shows the HL-GRB gamma-ray spectrum for $r={10}^{15}$
cm and $\Gamma = {10}^{3}$ where heavy nuclei can survive (see
Fig. 2). A thin dashed line shows the only cosmic-ray synchrotron
component from one HL GRB without attenuation by the CMB and CIB. 
A thick dotted line shows the LL-GRB gamma-ray spectrum for $r=6
\times {10}^{15}$ cm and $\Gamma = {10}$ where heavy nuclei can 
survive (see Fig. 3). A thin dotted line shows the only cosmic-ray synchrotron
component from one LL GRB without attenuation by the CMB and CIB.
Note that synchrotron self-inverse Compton components by accelerated 
electrons are not shown. We assume the comsic-ray composition with 
proton 75 \% and iron 25 \%. $\xi_{\rm{acc}} \equiv
U_{\rm{CR}}/U_{\gamma}$ and $\xi_{B} \equiv 
U_B/U_{\gamma}$ are set to 20 and 10 for the HL GRB case,
while 10 and 1 for the LL GRB case (see Appendix B). We use the low-IR
model for the CIB, which is presented by Kneiske et al. \cite{Kne1}. Fluence
sensitivity curves of GLAST and MAGIC are also shown \cite{GLA1,MAG1}.}
\end{figure}

Gamma rays can be produced by not only cosmic-ray synchrotron
radiation but also neutral pions, charged pions, muons and
electron-positron pairs generated via photomeson production. 
Gamma rays originating from photomeson production have very high
energies (e.g., $\gtrsim 10$ PeV for gamma rays coming from neutral pion
decay). Such gamma rays cannot avoid attenuation by the CIB, CMB and
cosmic radio background. In fact, the mean free path of 10 PeV photons
against pair-creation by the CMB photons is $\sim 10$ kpc, so that we
cannot expect to detect such gamma rays directly \cite{Der1}. 
Secondary electron-positron pairs generated by pair-creation are still 
energetic and up-scatter cosmic background photons. These boosted
photons can create pairs if they are still energetic, and the process 
repeats itself until the energy of degraded photons is in the $1-10$
TeV range \cite{Arm1,Asa3}. The mean free path of these
regenerated $1-10$ TeV photons is longer than 100 Mpc, and they can
reach the Earth. Whether the detection of these secondary gamma rays
is possible or not depends strongly on the intergalactic magnetic
field strength \cite{Com3}. As it is large, the expected
secondary flux becomes highly suppressed (see, e.g., \cite{Mur6} and
references there in). It is because the duration
of secondary emission becomes long and the emission becomes isotropic 
(for $\gtrsim {10}^{-16}$ G). Only when the intergalactic magnetic
field is very weak, which can be expected in the void region rather
than the structured region, the secondary emission can be detected. Even in
such cases, we expect that the secondary flux is sub-dominant compared
to the primary flux for parameter sets in Fig. 9. 
In fact, the expected fluence originating from gamma rays from
photomeson production can be estimated as $E_{\gamma}^2 \phi_{\gamma} \sim
{10}^{-7} \, \rm{erg} \rm{cm}^{-2}$ for HL GRB and 
$E_{\gamma}^2 \phi_{\gamma} \sim {10}^{-8} \, \rm{erg} \rm{cm}^{-2}$,
respectively \cite{Arm1,Asa3}. In addition, the secondary fluence from
cosmic-ray synchrotron emission (estimated by using the same
method presented in Ref. \cite{Mur6}) is smaller than the primary
one. Therefore, we omit spectra of secondary delayed emission in this
paper \cite{Com4}. Note that the diffuse gamma-ray background from GRBs
will be much smaller than the EGRET limit, which is expected to be 
$E_{\gamma}^2 \Phi_{\gamma} \lesssim {10}^{-8} \, \rm{GeV} \,
\rm{cm}^{-2} {s}^{-1} \rm{sr}^{-1}$ \cite{Mur6,Wax7}. 
So far we have considered the internal shock model. For other models,
see Appendixes D and E.

\section{Summary and Discussion}
In this paper, we have shown that not only protons but also heavy
nuclei can be accelerated up to ultra-high energies in both of HL GRBs
and LL GRBs. We exploit the internal shock model, (external) reverse 
shock model and forward shock model. We have also discussed various 
implications for neutrino, gamma-ray and UHECR astronomy. Especially, 
we have studied cosmic-ray acceleration in LL GRBs which may play 
an important role as high-energy cosmic-ray sources.

Let us summarize this paper below.
First, we have shown that UHECR nuclei can be produced in both of HL
GRBs and LL GRBs.

(A1) In the internal shock model, UHE protons and heavier 
nuclei can be produced in both of HL GRBs and LL GRBs. However, 
the allowed parameter range is limited. At smaller radii, most
of the UHE nuclei are depleted, and a significant fraction of 
the nonthermal cosmic-ray energy is transferred into neutrinos.
On the other hand, survival of UHE nuclei is possible 
only at large radii and/or for large Lorentz factors. Typically, 
relatively large radii $r \gtrsim {10}^{15}$ cm will be required.
For HL GRBs with $L_{b} \sim {10}^{51-52} \, \rm{erg} \rm{s}^{-1}$, 
both of protons and heavier nuclei can be 
accelerated up to ultra-high energies above ${10}^{20}$ eV.
For LL GRBs with $L_{b} \sim {10}^{46-47} \, \rm{erg} \rm{s}^{-1}$, 
heavy nuclei can be accelerated above ${10}^{20}$ eV, while 
proton acceleration above such high energies may be difficult. 
In addition, we have shown that the effect of 
synchrotron self-absorption suggested by Wang et al. \cite{Wan2} is 
not so important due to the cross section in the non-resonant region.

(A2) In the (external) reverse shock model, UHE nuclei can be produced
in both of HL GRBs and LL GRBs. Not only heavy nuclei but also 
protons achieve ultra-high energies above $\sim {10}^{20}$ eV.
Whether heavy nuclei can survive or not depends on various
parameters, and the sufficiently low photon density is 
required for survival. That is, small $E_{\rm{ej}}$, $\epsilon_{e}$, 
$\epsilon_{B}$ and $n$ are needed. Conversely, if 
values of these parameters are large enough, in which optical/infrared 
are often expected, survival of UHE nuclei is impossible.  
Note that UHE nuclei produced at internal shocks could be
photodisintegrated by photons generated at the reverse shock, when the
ejecta is in the thick ejecta regime. Hence, we could not expect
survival of UHE nuclei, when we see strong
optical/infrared flashes.    

(A3) In the (external) forward shock model, not only heavy nuclei but also
protons can achieve ultra high energies above $\sim {10}^{20}$ eV in
both of HL GRBs and LL GRBs. When the photon density
is sufficiently low, survival of UHE nuclei is possible.
The relevant parameters $f_{N \gamma}$ and maximum energy
$E_{N}^{\rm{max}}$ depend on time. In the ISM case, $f_{N \gamma}$ 
monotonically increases with time before the jet break occurs. On the 
other hand, in the wind-medium case, $f_{N \gamma}$ monotonically
decreases with time. We have shown that the jet-break effect can be 
important for survival of UHE nuclei, which was not pointed out
previously. After the jet break, the photon density decreases, so 
that survival of UHE nuclei becomes easier.

In the internal shock model and reverse shock model, we have to assume
that heavy nuclei are contained in relativistic outflows of GRBs.
It is natural to consider that only light elements can be synthesized
in such a highly relativistic outflows due to the high photon to
baryon ratio~\cite{pruet02,lemoine02}.
However, there may be possibility that such an outflow is contaminated
with heavy nuclei that are produced by explosive 
nucleosynthesis~\cite{nagataki03,maeda03,nagataki06,tominaga07}.
Also, such heavy nuclei can be entrained from the stellar surroundings due
to Kelvin-Helmholtz instabilities and/or oblique shocks \cite{Zha2}. 
Whether entrained heavy nuclei can survive or not is the issue that should
be carefully examined. 
On the other hand, in the forward shock model and 
hypernova model, heavy nuclei will be supplied by the swept material. 
They could come from the stellar wind of the progenitor star,
e.g., the Wolf-Rayet star.

(B) Neutrino astronomy is useful as a direct probe of cosmic-ray
acceleration, although the detection is not so easy. High-energy
neutrinos from GRBs have been predicted in various contexts. 
If detected, we can obtain very useful information
on cosmic-ray acceleration in GRBs. In this paper, we have calculated 
neutrino spectra from HL GRBs and LL GRBs in the internal shock model.
We have shown that the neutrino detection by near-future telescopes such
as IceCube and KM3Net would be difficult, when acceleration and
survival of UHE 
heavy nuclei (such as iron) are possible. In the internal shock model,
for example, we can expect $N_{\mu} \sim 1$ event from one GRB at $z=0.1$ and
$N_{\mu} \sim 50$ events/yr as the cumulative background for
one of the typical parameter sets (where UHE heavy nuclei cannot
survive), with large baryon loading factors required in the GRB-UHECR 
hypothesis. On the other hand, survival of UHE heavy nuclei leads to 
much smaller neutrino fluxes, as we have demonstrated. Hence, 
second generation neutrino telescopes would be needed for neutrino
detection in the latter case, although observations by
IceCube and KM3Net would be useful to constrain the GRB-UHECR hypothesis.

(C) We can also expect high-energy gamma-ray emissions (including leptonic and
hadronic gamma rays) from GRBs as well as high-energy neutrinos. 
We have shown that high-energy gamma rays above TeV can escape from the
source, when UHE heavy nuclei (such as iron) can survive. As an
example, we have calculated and evaluated gamma rays coming from accelerated  
cosmic rays in the internal shock model. If the source is nearby, we
have seen that high-energy gamma rays from cosmic-ray synchrotron
radiation would be detected by GLAST, MAGIC and VERITAS, with large 
baryon loading factors required in the GRB-UHECR
hypothesis. (However, we should keep in mind that the synchrotron
self-inverse Compton radiation by accelerated electrons could be more
important.) Hence, the high-energy gamma-ray detection will also be 
one of the important clues to testing the GRB-UHECR hypothesis. 
When the intrinsic pair-creation optical
depth becomes high enough (which infers UHE nuclei cannot survive), 
escape of TeV gamma rays becomes impossible and the intrinsic
pair-creation cutoff should exist. But we may still find
signatures of UHECR acceleration \cite{Asa2}, although we cannot
expect survival of UHE heavy nuclei in GRBs. 

In this paper, we have demonstrated spectra of high-energy neutrinos and 
gamma rays, which are expected if GRBs are UHECR accelerators, by
using detailed numerical calculations. We have also discussed 
detectabilities of such high-energy emission in the near future.    
Not only neutrino and gamma-ray astronomy but also UHECR astronomy are
becoming important to test the GRB-UHECR hypothesis. In Appendix F, we 
have discussed implications of the GRB-UHECR hypothesis for UHECR
astronomy by exploiting qualitative arguments. Future observations of 
UHECRs may allow us to distinguish between bursting sources and steady 
sources. If UHECR sources are bursting sources such as GRBs, we expect
to obtain information on the effective EGMF and source number
density by using the observed anisotropy of UHECRs. If we can, we 
could constrain the local bursting rate, for example, which may allow us to
distinguish between HL GRBs and LL GRBs.

When we are preparing this work, we found that 
Wang et al. \cite{Wan2} studied acceleration of UHE nuclei in GRBs
independently. 
In this paper, we have studied in more detail by making use of the
detailed cross section and spectra of afterglows, and calculate
spectra of high-energy neutrinos and gamma rays. 
Furthermore, we have also studied cosmic-ray acceleration in LL GRBs. 
     

\acknowledgments
KM thanks C. Dermer because he made very profitable comments at
the conference TAUP2007 on September, 2007.
KM also thanks the referee, R. Blandford, X.Y. Wang, Y.Z. Fan, 
K. Asano, S. Inoue and H. Takami. We thank R. Yamazaki and K. Toma for 
helpful comments. The work of KM is supported by a Grant-in-Aid for JSPS. 
The work of KI is supported by Grants-in-Aid for Scientific Research
of the Japanese Ministry of Education, Culture, Sports, Science,
and Technology 18740147 and 19047004. The work of SN is supported by 
Grants-in-Aid for Scientific Research of the Japanese Ministry of Education,
Culture, Sports, Science, and Technology 19104006, 19740139, 19047004.
The work of TN is supported by Grants-in-Aid for Scientific Research
of the Japanese Ministry of Education, Culture, Sports, Science,
and Technology 19047004. The work of us is supported by a Grant-in-Aid for the 
21st Century Center of Excellence ``Center for Diversity and University in 
Physics'' from the Ministry of Education, Culture, Sports, Science and 
Technology of Japan. 





\appendix
\section{Cosmic-Ray Acceleration at Internal Shocks of LL GRBs}
\begin{figure}[b]
\includegraphics[width=0.97\linewidth]{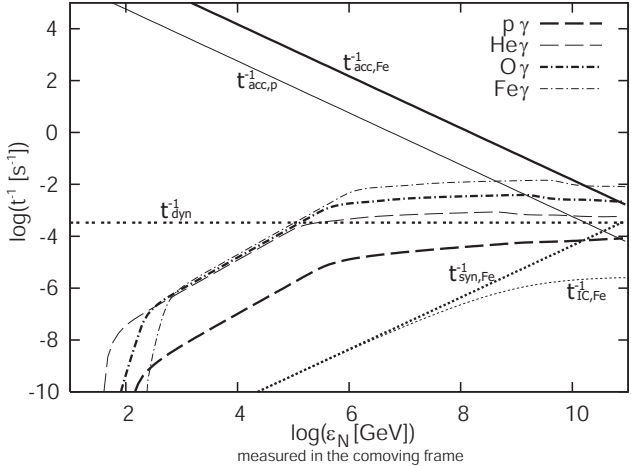}
\caption{\label{f10} The same as Fig. 3. But used parameters are 
$L_{b}={10}^{47} \, \rm{erg} \, {s}^{-1}$, $\varepsilon_{\rm{ob}}^b
 =5$ keV, $\Gamma=10$, $r=9 \times {10}^{14}$ cm (corresponding to 
$\delta t/(1+z)=150$ s) and $\xi_{B} (\approx \epsilon_B/\epsilon_e)=1$. 
Note that this parameter set is the same one as used in Ref. \cite{Mur2}
and it implies that a moderate fraction of the energy carried by
protons goes into neutrinos.}
\end{figure}
\begin{figure}[b]
\includegraphics[width=0.97\linewidth]{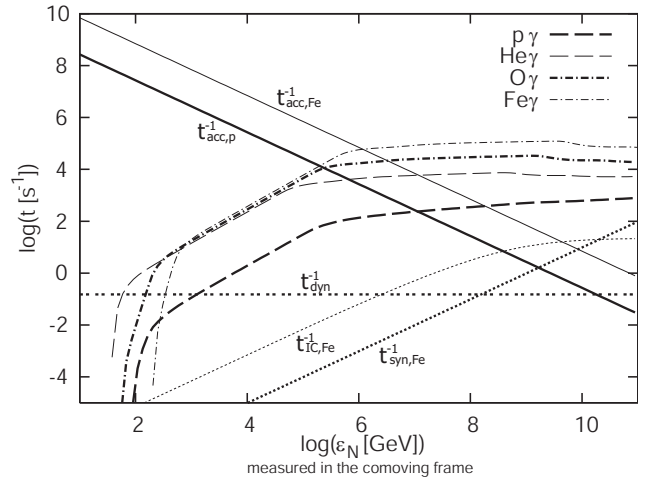}
\caption{\label{f11} The same as Fig. 3. But used parameters are 
$L_{b}={10}^{47} \, \rm{erg} \, {s}^{-1}$, $\varepsilon_{\rm{ob}}^b
=5$ keV, $\Gamma=5$, $r=7 \times {10}^{12}$ cm, $l={10}^{11}$ cm 
and $\xi_{B} (\approx \epsilon_B/\epsilon_e)=1.1 \times {10}^{-2}$. 
Note that this parameter set is the same one as used in Ref. \cite{Ghi1}
and it does not allow for accelerating cosmic rays up to ultra-high energies.}
\end{figure}

In this section, we discuss acceleration and survival of UHE nuclei in the
internal shock model of LL GRBs. We think that it is important to
demonstrate whether acceleration and survival of UHE nuclei are
possible or not, because a kind of fine tuning is necessary. In fact,
the parameter range such that acceleration and survival of UHE nuclei is
possible (see Fig. 3) is rather limited, and we can more easily find
parameter sets such that acceleration of survival of UHE nuclei is impossible.
Note that collision radii of LL GRBs are largely uncertain. We
can expect $r \sim {10}^{15-16}$ cm, which implies the time scale 
$\delta t \sim {10}^{2-3}$ s \cite{Mur2,Gup1,Tom1}. In fact, the observed
light curve of GRB 060218 is relatively simple and smooth. However, the 
collision radius may be much smaller as suggested in Ref. \cite{Ghi1}.
 
As in cases of HL GRBs, we can evaluate the maximum energies as (see
also Ref. \cite{Mur2})
\begin{eqnarray}
(1+z)E_{N,\rm{ad}}^{\rm{max}} &=& \frac{\Gamma ZeBl}{\eta} \nonumber \\
&\simeq& 2.2 \times {10}^{20} \, \mr{eV} \, Z \eta^{-1}
\epsilon _{B}^{1/2} \epsilon_e^{-1/2} \nonumber \\ &\times& {\left[ 
\frac{\Gamma_{\mr{sh}}(\Gamma_{\mr{sh}}-1)}{2} \right]}^{1/2}\!\!\!
L_{\gamma,47}^{1/2} {\Gamma}_{1}^{-1} \label{LLEad}
\end{eqnarray}
and
\begin{eqnarray}
(1+z)E_{N,\rm{syn}}^{\rm{max}} &=& \sqrt{ \frac{6 \pi Ze}
{Z^4 \sigma _{T} B \eta}}\frac{\Gamma m_{N}^2
c^2}{m_e} \nonumber \\
&\simeq& 7.4 \times {10}^{19} \, \mr{eV} \, A^2 Z^{-3/2} \eta^{-1/2}
\epsilon _{B}^{-1/4} \epsilon_e^{1/4} \nonumber \\ &\times& {\left[ 
\frac{\Gamma_{\mr{sh}}(\Gamma_{\mr{sh}}-1)}{2} \right]}^{-1/4} \!\!\!\!
L_{\gamma,47}^{-1/4} {\Gamma}_{1}^{3/2} r_{15}^{1/2}. \label{LLEsyn}
\end{eqnarray}
Therefore, we expect that LL GRBs also can produce UHECRs, especially 
that UHE nuclei can be accelerated above ${10}^{20}$ eV. 
These two conditions (Eqs. (\ref{LLEad}) and (\ref{LLEsyn}))
equivalently lead to inequalities (ignoring the term
$(1+z)$) as 
\begin{eqnarray}
0.5 Z^{-1} \eta \Gamma_1 E_{N,20} &\lesssim& 
L_{\mr{M},48}^{1/2} \epsilon_{B,-1}^{1/2} 
{\left( \frac{\Gamma _{\mr{sh}}(\Gamma_{\mr{sh}}-1)}{2} 
\right)}^{1/2} \nonumber \\
&\lesssim& 0.55 A^4 Z^{-3} {\eta}^{-1} r_{15} {\Gamma}_1^3 E_{N,20}^{-2}, \label{LLpro}
\end{eqnarray}
which is the same as Eq. (2) of Ref. \cite{Mur2}. 
As seen from the above inequalities, acceleration of protons above
$\sim {10}^{20}$ eV in LL GRBs will be more difficult than that in HL
GRBs, because necessary parameters should be suitably chosen. For
protons to be accelerated above $\sim {10}^{20}$ eV, relatively more 
luminous/magnetized LL-GRBs with higher Lorentz factors are required. 
In fact, GRB 031203, whose redshift
and luminosity are intermediate between two dim events (GRB 060218 and 
GRB 980425) and usual HL GRBs, might be such an event \cite{Gue2,Mal1}.  
Eqs. (\ref{LLEad}) and (\ref{LLEsyn}) just mean that UHECR production 
is also possible in relatively lower luminous GRBs such as XRFs and LL
GRBs, even if LL GRBs do not form a distinct population.

In Fig. 3, we have shown the results in cases where acceleration and
survival of UHE nuclei are possible. 
For comparison, we show the results for the parameter set used in 
Ref. \cite{Mur2}. In this case, the collision radius is smaller, so
that we cannot expect survival of UHE nuclei. Protons can be accelerated up to 
$\sim 5 \times {10}^{19}$ eV while Fe nuclei up to $\sim {10}^{20}$ eV
(but photodisintegrated).

In Fig. 5, we show the results for parameters inferred by Ghisellini et
al. \cite{Ghi1}. A collision radius is small enough, $r=7 \times
{10}^{12}$ cm, so that the photomeson production and
photodisintegration efficiencies are rather high.
Because the photodisintegration and photomeson
production processes prohibit acceleration of cosmic rays up to 
very high-energy, we cannot even expect UHECR production in this case.

\section{Energetics of GRBs and UHECRs}
In this section, let us briefly overview the energetics of GRBs and UHECRs in
order to see that GRBs (and hypernovae) could be one of the candidates 
of UHECRs (despite of some caveats). The detailed arguments are
found in Refs. \cite{Wax1,Mil1,Vie1,Wic1,Ber1,Der1,Scu1,Ber2,Vie3,Wax5}. 
In the previous section, we have seen that relativistic 
outflows that make GRBs satisfy various 
conditions for cosmic rays to be accelerated up to greater 
than ${10}^{20}$ eV. Here, let us compare the energy generation rate 
of observed UHECRs with the gamma-ray energy generation rate of GRBs. 
The UHECR generation rate given by Ref. \cite{Wax5} from Fly's Eye
data is $E_{\rm{CR}}^2 d \dot{N}_{\rm{CR}}/dE_{\rm{CR}} \approx 
0.65 \times {10}^{44} \, \rm{ergs} \, \rm{Mpc}^{-3} \rm{yr}^{-1}$. 
The UHECR generation rate from the recent PAO results \cite{Yam1} is 
compiled by Dermer \cite{Der1}. The UHECR generation rate at ${10}^{19}$ eV is
$E_{\rm{CR}}^2 d \dot{N}_{\rm{CR}}/dE_{\rm{CR}} \approx 0.8 \times {10}^{44}
\, \rm{ergs} \, \rm{Mpc}^{-3} \rm{yr}^{-1}$. At higher energies, the
value becomes smaller.

The radiation energy generation rate of GRBs is given by the product
of the local GRB rate $\rho(0)$ (without the beaming-correction) and
the average (isotropic equivalent) gamma-ray energy release
$E_{\gamma}^{\rm{iso}}$. First, let us consider HL GRBs. Recently, 
Kocevski \& Butler \cite{Koc1} provided $E_{\gamma,[1 \rm{keV}, 10 
\rm{MeV}]}^{\rm{iso}} \sim (4.1-7.8) \times {10}^{52}$ ergs, in a 
rest frame bandpass $(1-{10}^{4})$ keV. Emission at higher energies 
(above the $(1-{10}^{4})$ keV band) implies that the total 
gamma-ray energy input can be even higher. Assuming $\beta \sim 2$, we
can expect $E_{\gamma}^{\rm{iso}} \sim (1-2) \times {10}^{53}$
ergs. The energy input at high energies is rather uncertain, because
it depends on the high-energy photon index, pair-creation cutoff and
possible inverse-Compton contribution. Here, we just adopt $2.5 \times {10}^{53}$
ergs as just a reference value for HL GRBs.  
On the other hand, by using logN-logS relationship with the assumption that 
the GRB rate traces, e.g., the star formation rate, various 
authors estimated the local GRB rate. In the pre-Swift era, 
the estimated long GRB rate was $\sim (0.2-1) \, \rm{Gpc}^{-3} 
\rm{yr}^{-1}$ (see, e.g., \cite{Gue3} and references there in). 
After the launch of \textit{Swift}, many GRBs including high-redshift
bursts were observed, and various authors provided the local GRB rate
(see, e.g., \cite{Le1,Gue1}). For example, Ref. \cite{Gue1} obtained 
$(0.2-0.3) \, \rm{Gpc}^{-3} \rm{yr}^{-1}$ assuming that the GRB rate
traces the star formation rate. However, some authors claimed that the 
GRB rate has a faster evolution with redshift, which leads to the
lower local GRB rate, $\sim 0.05 \, \rm{Gpc}^{-3} \rm{yr}^{-1}$
\cite{Le1,Gue1,Kis1}.
Next, we shall consider LL GRBs. However, the radiation energy release 
and local rate of LL GRBs are much more uncertain. The radiation
energy of GRB 060218 is $\sim {10}^{50}$ ergs, while that of GRB
980425 is an order of magnitude smaller.  
The suggested local LL GRB rate is $\sim {10}^{2-3} \, \rm{Gpc}^{-3}
\rm{yr}^{-1}$ which are likely to be much higher than the local HL GRB rate 
\cite{Gue2,Lia1}. Note that we cannot exclude 
possibilities that there is no such a distinct population, although we
assume the existence of LL GRBs in this paper. Too large rates might 
also be impossible due to constraints from observations of SNe
\Roman{ichi}bc. For example, Soderberg et
al. \cite{Sod1} argued that at most $\sim 10 \, \%$ of SNe \Roman{ichi}bc are
associated with off-beam GRBs based on their late-time radio
observations of 68 local SNe \Roman{ichi}bc. 

Now, we can estimate the UHECR generation rate of GRBs from the
radiation energy generation rate with an unknown baryon loading
factor, which is defined by $U_{\rm{CR}} \equiv \xi_{\rm{acc}} U_{\gamma}$. 
Then, the cosmic-ray energy generation rate of GRBs can be written as
$\mathcal{E}_{\rm{CR}}^{\rm{iso}}= \xi_{\rm{acc}} 
E_{\gamma}^{\rm{iso}}= \xi_{\rm{acc}}
(4 \pi r^2 l) \Gamma U_{\gamma} N$. Here, let us introduce $R$ which
is defined as ${R}^{-1} \equiv (E_{\rm{CR}}^2\frac{d
N_{\rm{CR}}}{dE_{\rm{CR}}})/\mathcal{E}_{\rm{CR}}$, where
$E_{\rm{CR}}^2 \frac{d N_{\rm{CR}}}{dE_{\rm{CR}}}$ is the cosmic-ray 
energy input at ${10}^{18.5-19.5}$ eV. When we assume $p=2$ as the 
source spectral index of UHECRs, we have $R \equiv 
\ln(E_{\rm{CR}}^{\rm{max}}/E_{\rm{CR}}^{\rm{min}})$.
The minimum energy is not well determined theoretically, but we can expect 
$E_{N}^{\rm{min}} \sim$ a few$\times \Gamma m_N c^2 \sim
{10}^{11.5}$ eV. We have also assumed that 
cosmic rays can be accelerated up to ultra-high energies, so that we 
can take $E_{N}^{\rm{max}} \gtrsim {10}^{20}$ eV.  
Hence, for HL-GRBs, we obtain \cite{Typ2}
\begin{eqnarray}
E_{\rm{CR}}^2\frac{d \dot{N}_{\rm{CR}}}{dE_{\rm{CR}}} 
&=& 5.0 \times {10}^{43} \, {\mr{ergs}
\, \rm{Mpc}^{-3} \, \rm{yr}^{-1}} {\left(\frac{\xi _{\mr{acc}}}{20}
\right)}  {\left(\frac{20}{R} \right)} \nonumber \\
&\times& \left( \frac{E_{\gamma}^{\rm{iso}}}{2.5 \times {10}^{53}
\, \rm{ergs}} \right) \left( \frac{{\rho}_{\mr{HL}}(0)}
{0.2 \,  \mr{Gpc}^{-3} \, \mr{yr}^{-1}} \right). \label{budHL}
\end{eqnarray}
and for LL GRBs,
\begin{eqnarray}
E_{\rm{CR}}^2\frac{d\dot{N}_{\rm{CR}}}{dE_{\rm{CR}}} 
&=& 5.0 \times {10}^{43} \, {\rm{ergs}
\, \rm{Mpc}^{-3} \, {\rm{yr}}^{-1}} {\left(\frac{\xi _{\mr{acc}}}{10}
\right)}  {\left(\frac{20}{R} \right)} \nonumber \\
&\times& \left( \frac{E_{\gamma}^{\rm{iso}}}{2 \times {10}^{50}
\, \rm{ergs}} \right) \left( \frac{{\rho}_{\mr{LL}}(0)}
{500 \,  \mr{Gpc}^{-3} \, \mr{yr}^{-1}} \right). \label{budLL}
\end{eqnarray}
Therefore, the UHECR energy generation rate is roughly comparable to
the radiation energy generation rate of GRBs (unless, for example, we
use the smaller local GRB rate inferred from the faster evolution with
redshift). In other words, we can expect $E_{\rm{CR}}^2 
\frac{d\dot{N}_{\rm{CR}}}{dE_{\rm{CR}}}({10}^{19} \, {\rm{eV}}) \sim
\varepsilon_{\rm{ob}}^2 \frac{d \dot{N}_{\gamma}}{d
\varepsilon_{\rm{ob}}}(\varepsilon_{\rm{ob}}^b)$ for both of HL GRBs and
LL GRBs, and two populations can supply the necessary amount
of UHECRs, when $\xi_{\rm{acc}} \sim 10$ and $p \sim 2$. From
Eq. (\ref{budHL}), we may have the required baryon loading factor 
$\xi_{\rm{acc}} \gtrsim 20$ for HL GRBs, which is consistent
with Ref. \cite{Der1}. 
(If the local HL GRB rate is very low, as recently suggested by some
authors, very large baryon loading factors $\xi_{\rm{acc}} \gtrsim
100$ will be required, which would be implausible \cite{Mur1}).
However, its value is not so solid and smaller $\xi_{\rm{acc}}$ 
could be possible, because the local
GRB rate (and the total radiation energy input and the minimum cosmic-ray 
energy) is not much certain, and we may expect that GRB is one of the UHECR
candidates at present.  

In the reverse-forward and/or hypernova models, we usually use the kinetic
energy of ejecta instead of the radiation energy of the prompt emission.
The nonthermal cosmic-ray energy is written as $\mathcal{E}_{\rm{CR}}
\equiv \epsilon_{\rm{acc}} E_{\rm{ej}}$. In the case of supernova, 
$\epsilon_{\rm{acc}} \gtrsim 0.1$ is typically needed to explain the cosmic-ray
flux below the knee. From observations of afterglows, we typically have
$E_{\rm{ej}} \sim {10}^{52-53}$ ergs for HL GRBs, while  
$E_{\rm{ej}} \sim {10}^{50-51}$ ergs for LL GRBs.
In the hypernova model, the total kinetic energy of the hypernova ejecta, 
$E_{\rm{ej}} \sim 5 \times {10}^{52}$ ergs is expected (see
\cite{Wan1} and references there in). But note that UHECRs can be 
produced only in the high velocity ejecta
with $\Gamma \beta \gtrsim 0.5$, which carries $E_{\rm{ej}} \sim 2 
\times {10}^{51}$ ergs \cite{Wan1,Wan2}.
For these models, instead of Eq. (\ref{budLL}), we have 
\begin{eqnarray}
E_{\rm{CR}}^2\frac{d\dot{N}_{\rm{CR}}}{dE_{\rm{CR}}} 
&=& 4.0 \times {10}^{43} \, {\rm{ergs}
\, \rm{Mpc}^{-3} \, {\rm{yr}}^{-1}} \, \epsilon_{\rm{acc}}
{\left(\frac{25}{R} \right)} \nonumber \\
&\times& \left( \frac{E_{\rm{ej}}^{\rm{iso}}}{2 \times 
{10}^{51} \, \rm{ergs}} \right) \left( \frac{{\rho}_{\mr{LL}}(0)}
{500 \,  \mr{Gpc}^{-3} \, \mr{yr}^{-1}} \right). \label{budAG}
\end{eqnarray}
Hence, UHECRs could be explained in the external shock model and/or 
hypernova model for LL GRBs as well as the external shock model for 
HL GRBs. A significant fraction of the kinetic energy must be 
injected to the nonthermal cosmic-ray energy.
As in the case of
supernovae, $\epsilon_{\rm{acc}} \sim 0.1-1$ will be required. 
Note that the hypernova model is originally considered
in order to explain high-energy cosmic rays above the second 
knee, where the cosmic-ray spectrum is formed by the superposition of 
cosmic-rays from hypernova ejecta with various velocities. 
The velocity profile with $d E_{\rm{ej}}/d (\Gamma \beta) \propto 
{(\Gamma \beta)}^{-3}$ leads to the superposed spectral index $p \sim 3$
(which seems too steeper than the typical values of the required
source spectrum index around the second knee for protons, $p \sim 2.4-2.7$).
In order to explain UHECRs above the ankle only by the hypernova
model, we would need some additional reason (which was not given in
Refs. \cite{Wan1,Wan2}), such as, e.g., the variable index of the velocity
profile or the velocity-dependent baryon loading. Instead, 
the hybrid model (GRBs and hypernovae) might be possible.

The above estimates are based on the assumption that the source spectral
index is $p=2.0$. However, if it is steeper, the required energy
generation rate becomes much larger. In the dip model suggested by
\cite{Ber1,Alo1}, the typical spectral index is $p \sim 2.4-2.7$,
which is steeper than $p \sim 2.0-2.3$ in the ankle model.
Such a steep spectrum leads to the larger baryon loading factor,
unless we change $E_{\rm{CR}}^{\rm{min}}$. For example, we have 
$R({10}^{19} \, {\rm{eV}}) \equiv {\mathcal{E}}_{\rm{CR}}
/(E_{\rm{CR}}^2 \frac{dN_{\rm{CR}}}{dE_{\rm{CR}}}({10}^{19} \rm{eV}))
\sim 2500$ for $p=2.4$ and $E_{\rm{CR}}^{\rm{min}} \sim {10}^{11.5}$ eV, 
which can be obtained easily from its definition.
Furthermore, we have neglected the cosmic-ray energy loss at the source. 
For example, the efficient neutrino production leads to depletion of 
high-energy cosmic rays if the shock dissipation at sufficiently inner radii.
When $f_{p\gamma}$ is not small, the required nonthermal cosmic-ray
energy will be raised by $\sim 1/(1-f_{p\gamma})$. In addition, the 
adiabatic loss and escape of particles should be taken into account
for more realistic calculations.

From Eqs. (\ref{budHL}), (\ref{budLL}) and (\ref{budAG}), we expect
that GRBs can be the main UHECR sources in both of the internal shock
model, and the external reverse and forward shock model. However,
note that too large nonthermal baryon loading factors seem implausible because 
the available energy for nonthermal cosmic rays would be limited,
e.g., by gravitational energy of falling 
materials of massive stars. In addition, the high radiative efficiency
of the prompt emission \cite{Iok1} might infer that the large baryon loading is
impossible, as long as the nonthermal baryon energy is smaller than the
thermal baryon energy. Although these possible caveats, current theories
cannot answer whether efficient UHECR production in GRBs is possible
or not. Rather, we should test this GRB-hypothesis from observations. 
Current and feature observations of UHECRs, neutrinos and gamma rays will
give us useful information. For example, observations by AMANDA and
IceCube have enabled us to constrain the averaged value of 
$f_{p \gamma}$ under the GRB-UHECR hypothesis \cite{Ach2,Ach3}. 

\section{The Neutrino Background}
It is very important to consider the neutrino background from GRBs
since it is not easy to detect neutrino signals from one GRB event.
The cumulative neutrino background flux from HL GRBs can be 
approximately evaluated by 
the following analytical expression using $R \sim 20$ \cite{Mur3,Wax7},
\begin{eqnarray}
E_{\nu}^2\Phi _{\nu} &\sim& \frac{c}{4\pi H_{0}}
 \frac{1}{4} \mr{min}[1,f_{p\gamma}] E_{p}^2 \frac{dN_{p}^{\rm{iso}}}{dE_{p}}
 {\rho}_{\mr{HL}}(0) f_{z} \nonumber\\
&\simeq& 4 \times 10^{-9} \mr{GeV cm^{-2} s^{-1} str^{-1}} \, 
\left( \frac{\xi _{\mr{acc}}}{20} \right) E_{\gamma,53}^{\rm{iso}} \nonumber\\
&\times& \left(\frac{f_{p\gamma}}{0.3}\right) \left(
\frac{{\rho}_{\mr{HL}}(0)}{0.2 \,
 \mr{Gpc}^{-3}\mr{yr}^{-1}}\right) 
\left( \frac{f_{z}}{3} \right), \label{nubkgHL}
\end{eqnarray}
and for LL GRBs we have \cite{Mur2}
\begin{eqnarray}
E_{\nu}^2\Phi _{\nu} &\sim& \frac{c}{4\pi H_{0}}
 \frac{1}{4} \mr{min}[1,f_{p\gamma}] E_{p}^2 \frac{dN_{p}^{\rm{iso}}}{dE_{p}}
 {\rho}_{\mr{LL}}(0) f_{z} \nonumber\\
&\simeq& 7 \times 10^{-10} \mr{GeV cm^{-2} s^{-1} str^{-1}} \, 
\left( \frac{\xi _{\mr{acc}}}{10} \right) E_{\gamma,50}^{\rm{iso}} \nonumber\\
&\times& \left(\frac{f_{p\gamma}}{0.05}\right) \left(
\frac{{\rho}_{\mr{LL}}(0)}{500 \,
 \mr{Gpc}^{-3}\mr{yr}^{-1}}\right) 
\left( \frac{f_{z}}{3} \right), \label{nubkgLL}
\end{eqnarray}
where $f_{z}$ is the correction factor for the possible contribution from
high redshift sources, which depends on the cosmology. In this paper,
we use the standard $\Lambda$CDM cosmology with 
$\Omega _{\mr{m}}=0.3, \Omega _{\Lambda}=0.7; H_{0}=71 \, \mr{km
\, s^{-1} \, Mpc^{-1}}$.

In general, we have to care about other backgrounds such as
the atmospheric neutrino background when we
consider the neutrino background. Observations of neutrinos from GRBs
have merits that we can take the time- and positional-coincidence.
Since HL GRBs are bright in the gamma-ray energy range, we expect the 
coincidence between neutrinos and gamma rays, which enables us to 
neglect the atmospheric and cosmogenic neutrino background essentially.
On the other hand, neutrino signals from LL-GRBs are very dim in the sense 
that most of the neutrino signals will not correlate with photon signals. 
Only for very nearby bursts, we might be able to expect such
correlations, and that it requires many-years operations.
However, we may see neutrino events that are positionally correlated
with SNe \Roman{ichi}c associated with LL GRBs. The angular resolution
of IceCube for neutrinos is about $1$ degree or so, which might be
searched by the optical-infrared follow-ups with ground-based
optical telescopes \cite{Mur2}. Of course, it is necessary that 
the LL GRB neutrino background makes the dominant contribution to the
neutrino background in this case.

Note that the GRB-UHECR hypothesis might lead to the enhanced
cosmogenic neutrino background. Cosmogenic neutrinos are generated
when UHECRs propagate in the universe, and they are the most promising 
very high-energy neutrino signals. Recent studies
have suggested that GRBs may have the strong evolution with
redshifts, so that UHECR sources have the strong evolution if they are GRBs.
In general, the strong evolution of UHECR sources leads to the higher
cosmogenic neutrino background flux, which may be useful as one of the
indirect clues to the GRB-UHECR hypothesis \cite{Yuk1}.
In addition, these neutrino signals can give us information on the
source spectral index, that is, they are useful to distinguish between
the ankle model and the dip model \cite{Tak4,All2}.    

\section{High-Energy Neutrinos and Gamma Rays in the Reverse-Forward
Shock Model}
In the reverse-forward shock model as well as the internal shock
model, neutrino signals are expected. 
The neutrino emission in the early afterglow phase was studied in
detail in Refs. \cite{Mur4,Der4}. We can evaluate the photomeson
production efficiency as \cite{Mur4}
\begin{equation}
f_{p\gamma} \simeq 0.088 \frac{L_{b,48}}{r_{16} {\Gamma_{2}}^2
{\varepsilon}_{\rm{ob},10 \, \mr{eV}}^{b}} \left\{ \begin{array}{rl} 
{(E_p/E_p^b)}^{\beta-1} & \mbox{($E_p < E_{p}^{b}$)}\\
{(E_{p}/E_{p}^{b})}^{\alpha-1} & \mbox{($E _p^{b} < E_p$)} 
\end{array} \right. \label{pgAG}
\end{equation}
where $\varepsilon_{\rm{ob}}^b$ is $\varepsilon_{\rm{ob}}^m$
or $\varepsilon_{\rm{ob}}^c$ or $\varepsilon_{\rm{ob}}^{sa}$.
In the reverse-forward shock model, we typically obtain $f_{p \gamma}
< 1$ except in the highest energies. The typical neutrino energy is around 
$\sim {10}^{18}$ eV rather than $\sim {10}^{15}$ eV (which is expected
for the prompt emission). Hence, expected muon event rates by IceCube
are generally small, and the detection is not so easy even by other
detectors such as PAO. Nevertheless, since EeV neutrinos are 
produced by $\sim {10}^{20}$ eV protons, the detection of such very
high-energy neutrinos is useful for diagnosing UHECR acceleration
at acceleration sites. (For the prompt emission, the typical
neutrino energy is $\sim$PeV, and PeV neutrinos are produced by
protons with $\sim {10}^{17}$ eV, smaller than ultra high energies.)  

Next, let us consider the forward shock model. From Eq. (\ref{pgAG})
and various quantities derived under the forward shock model, we can
obtain the photomeson production efficiency. By replacing 
$\varepsilon^b$ and $L_b$ with $\varepsilon_{\rm{ob}}^c$ and $\Gamma^2 
{(\varepsilon^c \varepsilon^m)}^{1/2} L_{\varepsilon,\rm{max}}$, respectively, 
we have
\begin{equation}
f_{p \gamma} \simeq 2.1 \times {10}^{-2} g_{-1}
\epsilon_{e,-1} \epsilon_{B,-1}^{3/2} E_{\rm{ej},53} n_0^{3/2} 
\left\{ \begin{array}{ll} {(E_p/E_p^b)}^{1} \\
{(E_{p}/E_{p}^{b})}^{0.5}
\end{array} \right.
\end{equation}
As seen in the previous section, $f_{p \gamma}$ increases with time
in the ISM case (before the jet break), while decreases in the
wind-medium case. For example, when we assume $\epsilon_{\rm{acc}}
E_{\rm{ej}}={10}^{54}$ ergs and $\rho_{\rm{HL}}(0)=0.2 \,
\rm{Gpc}^{-3} \rm{yr}^{-1}$, the expected diffuse neutrino flux at 
the break energy $E_{\nu}^b \approx 0.05 E_{p}^{b}$ is 
$E_{\nu}^{2} \Phi_{\nu} \sim 6 \times {10}^{-11} g_{-1}
\epsilon_{e,-1} \epsilon_{B,-1}^{3/2} E_{\rm{ej},53} n_0^{3/2} (f_z/3) 
\, \rm{GeV} \rm{cm}^{-2} \rm{s}^{-1} \rm{sr}^{-1}$. In the higher
energies, $E_{\nu}^{2} \Phi_{\nu}$ becomes larger.

In the cases where UHE nuclei can survive as demonstrated in Figs. 6 
and 7, we have $f_{\rm{Fe} \gamma} \lesssim 1$ at 
$E_{\rm{Fe}} \sim {10}^{20}$ eV. From Eq. (\ref{pgandNg}), we obtain 
$f_{p \gamma} \lesssim 4 \times {10}^{-3}$ at $E_{p} \sim {10}^{20}$
eV. Hence, we can estimate the cumulative background neutrino flux as 
$E_{\nu}^2 \Phi_{\nu} \lesssim {10}^{-10}\, 
\rm{GeV} \rm{cm}^{-2} \rm{s}^{-1} \rm{sr}^{-1}$ at 
$E_{\nu} \sim {10}^{18}$ eV under the GRB-UHECR hypothesis.   

High-energy gamma-ray emission from the reverse-forward shock
model is expected as well as the internal shock model (see, e.g.,
\cite{Zha3,Fan2}).  
For example, let us consider the forward shock model. 
In deriving Eq. (\ref{ggandNg}), let us replace $\varepsilon^b$, 
$\alpha=1$ and $\beta=2.2$ with $\varepsilon^c$, 
$\alpha=1.5$ and $\beta=2$ (for $p=2$).  
As a result, the optical thickness for pair-creation at 
$\varepsilon_{\rm{ob}}={10}^{12} \, {\rm{eV}} \, 
\varepsilon_{\rm{ob},12}$ can be evaluated as 
\begin{eqnarray}
\frac{f_{\gamma \gamma} ({\varepsilon}_{\rm{ob}}={10}^{12}\, \rm{eV} )}
{ f_{N \gamma}(E_N={10}^{20} \, \rm{eV})} 
\simeq 0.025 {\left( \frac{A}{56} \right)}^{-0.815} 
{\left( \frac{\varepsilon^2_{\rm{ob},12} \varepsilon^c_{\rm{ob},2}}
{E_{N,20} \Gamma_2^2} \right)}^{0.5},
\end{eqnarray}
where we have assumed $\varepsilon_{\rm{ob}} \lesssim 
\tilde{\varepsilon}_{\rm{ob}}^c = \Gamma^2 {(m_e c^2)}^2 
/ \varepsilon_{\rm{ob}}^c$ and $E_N > E_N^b$, which can be realized at 
$t \sim {10}^{2}$ s. At later time, we can expect
$\varepsilon_{\rm{ob}} \sim {10}^{12} \, {\rm{eV}} \gtrsim 
\tilde{\varepsilon}_{\rm{ob}}^c$ and have 
\begin{eqnarray}
\frac{f_{\gamma \gamma} ({\varepsilon}_{\rm{ob}}={10}^{12}\, \rm{eV} )}
{ f_{N \gamma}(E_N={10}^{20} \, \rm{eV})} 
\simeq 0.27 {\left( \frac{A}{56} \right)}^{-0.815} 
{\left( \frac{\varepsilon_{\rm{ob},12}}{E_{N,20}} \right)}^{0.5}.
\end{eqnarray}
Hence, we can expect that $\sim$TeV gamma rays escape from the source
when UHE irons can survive. In fact, we typically expect such
cases in the afterglow phase \cite{Zha3,Fan2}.
Knowing the possible intrinsic pair-creation cutoff in
the afterglow emission might give us clues to whether UHE heavy
nuclei can survive or not. However, it may be difficult to
distinguish the intrinsic pair-creation cutoff from other
possibilities such as that due to 
attenuation by CMB/CIB photons and that due to the maximum accelerated
energy of electrons.   

\section{High-Energy Neutrinos and Gamma rays in the Hypernova Model}
In the hypernova model, not only nonthermal photons radiated from
relativistic electrons but also thermal photons (in the optical band)
exist at the relatively early time. High-energy neutrino emission
generated via photomeson production is expected, but the detection
would be difficult \cite{Wan1}. Instead, high-energy gamma rays may be
detected by GLAST and/or MAGIC in the future. 
We can show that the optical depth against these thermal photons
will be smaller than the unity. Although photon spectra of hypernovae are
very complicated, let us make the very simple estimation by
using a black-body spectrum. We have 
\begin{equation}
f_{\gamma \gamma} (\varepsilon_{\rm{ob}}) 
\sim 0.35 \left( \frac{k T}{1 \, \rm{eV}}\right)  \left( \frac{r}{1 \, 
\rm{pc}}\right) {\left( \frac{{\varepsilon}_{\rm{ob}}}{1 \, \rm{PeV}}
\right)}^{-2}. 
\end{equation}
Therefore, hadronic gamma rays from high-energy neutral pions, muons and pairs
above $\sim$PeV could escape from the source. They will be cascaded by 
CMB/CIB photons. After our work was public, Asano \& M\'esz\'aros
performed detailed calculations, and they also showed that high-energy gamma
rays can escape from the source \cite{Asa3}. 
They demonstrated that not only gamma rays
from cosmic-ray synchrotron radiation but also secondary gamma rays
generated via pair-creation by the CMB/CIB photons can be detected in
the future, if the intergalactic magnetic field is weak enough.

\section{GRBs and UHECR Astronomy}
The discussions toward UHECR astronomy have been recently begun. The
direct correlation between UHECR sources and observed UHECR arrival
directions is one of the approaches to study the source properties. 
This can be expected only if the deflection angle due to the galactic
magnetic field (GMF) and EGMF is small enough. The latter is poorly
known both theoretically and observationally. $\sim (40-95) \%$ of volume within
$100$ Mpc can be regarded as the void region, whose uniform magnetic
field is not well-known. The Faraday rotation measurement of radio
signals from distant quasars implies $B_{\rm{EG}}^{\rm{void}}
\lambda_{\rm{Mpc}}^{1/2} \lesssim {10}^{-9} \, \rm{G}
\rm{Mpc}^{1/2}$ \cite{Kro1,Val1}. In addition, there is also the 
structured EGMF which traces the local matter
distribution. Observationally, clusters of 
galaxies have strong magnetic fields with $(0.1-1) \, \mu$G at its
center. This structured EGMF is very important for UHECR propagation 
(see, e.g., \cite{Dol1,Tak1} and references there in).
The GMF is relatively well measured, but the magnetic field of the 
galactic halo is not known well. This GMF also affects trajectories of
UHECRs \cite{Tak2}. If UHECR trajectories are deflected only weakly, 
the deflection angle by EGMF can be written as
\begin{eqnarray}
\theta_d &\approx& \frac{\sqrt{2} Z e B_{\rm{EG}} D}{3 E_N
\sqrt{D/\lambda}} \nonumber \\ &\simeq& {2.5}^{\circ
}  Z_{1} E_{N,20}^{-1} B_{\rm{EG}.-10} 
\lambda_{\rm{Mpc}}^{1/2} D_{100 \, \rm{Mpc}}^{1/2} \label{DefRan} \\
\theta_d &\approx& \frac{Z e B_{\rm{EG}} D}{2 E_N} \nonumber \\ 
&\simeq& {2.6}^{\circ} Z_1 E_{N,20}^{-1} B_{\rm{EG},-11} D_{100 \,
\rm{Mpc}}. \label{DefCoh}
\end{eqnarray}
Eq. (\ref{DefCoh}) is for the coherent EGMF ($\theta_d D \ll \lambda$)
, while Eq. (\ref{DefRan}) is for the EGMF with the coherent length
$\lambda$ ($\theta_d D \gg \lambda$). 
Recent PAO results imply the $\sim 3^{\circ}$ deflection angle, 
assuming that nearby AGN are UHECR sources \cite{PAO1,PAO2}. 
The inferred value $B_{\rm{EG}} \sim {10}^{-10}$ G is consistent with 
the predicted EGMF strength, and it could
be used to set lower limits on the effective EGMF \cite{Der1}. 
However, in fact, 
we cannot neglect the effect of the GMF. Although the deflection angle
due to the GMF depends on the model of the galactic magnetic field, it
can become $\gtrsim 3^{\circ}$. Therefore, lower limits could be set only
if we see the direction where the effect of the GMF is small. 
In addition, the inferred value is derived from the positional
correlation between the arrival directions of UHECRs and the
incomplete AGN catalogue used by the PAO collaboration. Therefore, in general, 
more observations and careful examinations are required 
in order to estimate the effective EGMF from correlation signals. 

Although the GMF can be important for the deflection
angle, it does not affect the delayed time so much.
From Eqs. (\ref{DefRan}) and (\ref{DefCoh}), we can estimate the
delayed time of UHECRs as
\begin{eqnarray}
\tau_d &\approx& \frac{D \theta_d^2}{4 c} \simeq 
{ 10}^{3} \, {\rm{yrs}} \, Z^2 E_{N,20}^{-2} \nonumber \\
&\times& B_{\rm{EG},-10}^2 \lambda_{\rm{Mpc}} D_{100 \, \rm{Mpc}}^{2} 
\label{DelRan}\\
\tau_d &\approx&  \frac{D \theta_d^2}{2c} \simeq 
3.5 \times {10}^{3} \, {\rm{yrs}} \, Z^2 E_{N,20}^{-2} \nonumber \\
&\times& B_{\rm{EG},-11}^2 D_{100 \, \rm{Mpc}}^{3}. \label{DelCoh}
\end{eqnarray}
Eq. (\ref{DelCoh}) is for the coherent EGMF, while Eq. (\ref{DelRan})
is for the EGMF with coherent length $\lambda$. Later we shall use 
the former expression for convenience. Note that the EGMF in the above
equations expresses the effective EGMF field. 
UHE protons suffer from the stochastic energy loss due to photomeson
production above the GZK energy, so that the dispersion of the arrival 
time is also $\sim \tau_d$. As long as we can use Eq. (\ref{DelRan}), 
we also expect multiplets such that the lower energy cosmic-ray event
precede higher energy cosmic-ray event, because the time delay is 
statistically distributed \cite{Vie3}.  

From the delayed time and observed local GRB rate, we can estimate 
the number density of GRB-UHECR sources. Let us perform a simple 
analytic calculation for the number density of GRB-UHECR sources.
Following Ref. \cite{Mir1}, we adopt the top-hat model, where the effect of
energy losses are approximately negligible when $D<D_c(E_{\rm{CR}})$, 
and eliminating all cosmic rays coming from $D>D_c(E_{\rm{CR}})$. 
Here, $D_c(E_{\rm{CR}})$ is the effective cutoff distance.  
For example, we obtain $D_c (E_p={10}^{20} \, \rm{eV}) \approx 50$ Mpc 
when we define that $D_c (E_{\rm{CR}})$ is the radius from
which $1/e$ of cosmic rays can reach us (see, e.g., \cite{Har1}).
By assuming that we can observe GRB-UHECR sources only during the duration
time which is comparable to the delayed time $\sim \tau_d$, we have
\begin{equation}
F_{\rm{CR}} \approx \frac{1}{4 \pi D^2 \tau_d} 
\frac{dN_{\rm{CR}}}{dE_{\rm{CR}}}, \label{CRflux}
\end{equation}
where $F_{\rm{CR}}$ is the observed UHECR flux at a given energy.
Most cosmic rays above the effective cutoff radius $D_c$ cannot
reach us. Hence, the minimum flux $F_{\rm{CR}}^c$ for a given 
$\frac{dN_{\rm{CR}}}{dE_{\rm{CR}}}$ is obtained by substituting 
$D_{c}$ into $D$. We also have the corresponding delayed time 
$\tau_d^c$. The number density of GRB-UHECR sources at UHE energies 
is calculated by integrating
the differential source number density (the number of GRBs per unit 
observed UHECR flux) over the UHECR flux $F_{\rm{CR}} \gtrsim 
F_{\rm{CR}}^c$. As a result, we have $n_s \approx (3/5) 
\rho(0) \tau_d^c$. Note that, in the GRB-UHECR hypothesis, the 
number density of UHECR sources will depend on the observed energy
because $\tau_d^c$ depends on the energy. The number density of GRB-UHECR
sources observed at $E_{p}={10}^{19.9}$ eV becomes     
\begin{eqnarray}
n_s^{\rm{HL}} \approx \frac{3}{5} {\rho}_{\rm{HL}}(0) \tau_d^c 
&\simeq& 1.7 \times {10}^{-5} \, {\rm{Mpc}}^{-3} \, B_{\rm{EG},-9}^{2} 
\lambda_{\rm{Mpc}}\nonumber \\ 
&\times& \left( \frac{{\rho}_{\mr{HL}}(0)}{0.2 \,  \mr{Gpc}^{-3} \,
\mr{yr}^{-1}} \right) \label{nsHL} \\
n_s^{\rm{LL}} \approx \frac{3}{5} {\rho}_{\rm{LL}}(0) \tau_d^c
&\simeq& 4.2 \times {10}^{-5}  \, {\rm{Mpc}}^{-3} \, 
B_{\rm{EG},-10}^2 \lambda_{\rm{100 \, \rm{kpc}}} \nonumber \\ 
&\times&  \left( \frac{{\rho}_{\mr{LL}}(0)}{500 \,  \mr{Gpc}^{-3} \,
\mr{yr}^{-1}} \right). \label{nsLL}
\end{eqnarray} 
The number density of UHECR sources for arbitrary energies can be obtained
by calculating $D_c^2/E_{\rm{CR}}^2$. 
Eqs. (\ref{nsHL}) and (\ref{nsLL}) suggest that determination of the
number density of UHECR sources and effective EGMF enable us to test the
GRB-UHECR hypothesis. 
Acquiring the latter is especially not easy, hence one of the ultimate goals of
the UHECR astronomy. But future observations of UHECRs by PAO and TA will
give us clues to them via observations of correlation signals.

The source number density can be estimated from the study
of the small-scale anisotropy. If the observed small-scale 
clustering originates in the same sources, we can estimate the minimum
number density of UHECR sources \cite{Dub1,Har2}. For example, recent PAO
observations showed 6 pairs with separation smaller than the
correlation angle scale of $6^{\circ}$ among the 27 highest-energy events. 
This leads to a lower limit for the number of sources $\geq 61$
\cite{PAO2}. When we adopt $D_c \approx 130$ Mpc, we can estimate the 
source number density as $n_s \gtrsim 3 \times {10}^{-5} \, \rm{Mpc}^{-3}$.
In more detail, the significance of small-scale 
clustering in the arrival directions of UHECRs can be studied by the angular
two-point auto-correlation function. By simulating the arrival distribution
of UHECRs, the source number density that can reproduce observational
results can be estimated \cite{Kac1}. The constraint obtained from 
the small-scale anisotropy observed by AGASA is $n_s \gtrsim {10}^{-5} \,
\rm{Mpc}^{-3}$ for uniformly distributed proton sources. Despite large
uncertainties due to a small number of observed events, PAO
observations in the future can lead to the more robust estimation \cite{Tak3}. 

For example, if we can find that the source number density is 
$n_s \sim {10}^{-4} \, \rm{Mpc}^{-3}$, the required effective EGMF for a given
local GRB rate is $B_{\rm{EG},-9}^2 \lambda_{\rm{\rm{Mpc}}} \sim 10$
(which might be larger the upper limit of the uniform EGMF) for HL GRBs and 
$B_{\rm{EG},-9}^2 \lambda_{\rm{\rm{Mpc}}} \sim 0.005$ for LL GRBs.
However, since the robust estimation of the source number density may
be difficult, we may have only the lower limit. In such cases, for example, 
$n_{s} \gtrsim {10}^{-5} \, \rm{Mpc}^{-3}$ would infer the required EGMF 
$B_{\rm{EG},-9}^2 \lambda_{\rm{\rm{Mpc}}} \gtrsim 1$ for HL GRBs and
$B_{\rm{EG},-9}^2 \lambda_{\rm{\rm{Mpc}}} \gtrsim 0.0005$ for LL
GRBs. Therefore, determination of the upper limit on the effective EGMF is
important in order to test the GRB-UHECR hypothesis.
If we know that the effective EGMF is too weak, possibilities of
HL GRBs as observed main UHECR sources could be excluded. (However, we 
could still expect to see HL GRBs as rare but very
bright UHECR sources. In such cases, one HL GRBs could be found as a very 
bright event among $\sim \rho_{\rm{LL}}(0)/\rho_{\rm{HL}}(0) \sim 1000$ 
LLGRB-UHECR sources.) On the other hand, when the effective EGMF is strong
enough, HL GRBs rather than LL GRBs would be more plausible main UHECR 
sources. It is
because the too large source number density cannot explain the anisotropy 
\cite{Tak3}. Of course, when LL GRBs can also produce UHECRs, they could be
regarded as numerous faint sources, and it is difficult to exclude the 
existence of such faint sources. More generally, we should employ the 
luminosity-weighted distribution rather than the luminosity-uniform 
distribution. Even if LL GRBs do not form a distinct population, it
would be important to know the local GRB rate as a function of 
$E_{\gamma}^{\rm{iso}}$ for more refined investigations.  

Although Eqs. (\ref{nsHL}) and (\ref{nsLL}) suggest that future 
observations enable us to test the GRB-UHECR hypothesis, the 
above estimation is simple and not quantitative. In order to
present realistic arguments, we need the detailed simulation of 
cosmic-ray propagation, which is being planned as our future work. 
Note that, when UHECRs contain heavier nuclei, $\tau_d^c$ for heavy
nuclei differs from that for protons, so that the source number
density also depends on the composition of UHECRs. 

Before finishing this section, let us briefly describe observable 
features which are expected for GRB-UHECR sources.
In the GRB-UHECR model, we can expect several prominent features different from
those in the steady source models.
One can be observed in the clustering properties. Since the UHECR flux is
proportional to $D^{-4}$ as seen in Eq. (\ref{CRflux}), the differential
source number density is proportional to $F_{\rm{CR}}^{-9/4}$
as long as we assume the luminosity-uniform source distribution
\cite{Mir1}. This leads to the source number density above 
$F_{\rm{CR}}$ is proportional to $F_{\rm{CR}}^{-5/4}$, which 
differs from the steady source model where we have 
$\propto F_{\rm{CR}}^{-3/2}$. This difference leads to
the difference in the average number of multiplets, i.e., the difference in
the ratio of the expected number of clusters with different
multiplicities \cite{Har2}. In the GRB-UHECR model, we may obtain the ratio
between the number of triplets or higher multiplets and doublets,
which is $\sim 0.6$ in the GRB-UHECR model, while $\sim 0.33$ in the 
steady source model. 
 
Another prominent feature is the existence of the critical energy \cite{Mir1}.
As seen in Eqs. (\ref{nsHL}) and (\ref{nsLL}), the source number
density decreases with the energy. Since $D_c$ decreases rapidly above
the pion production threshold energy, $\tau_d^c$ will also show the
rapid decrease above the pion threshold energy. Therefore, there
exists the critical energy where the observed source number becomes
the unity. Around this critical energy, we have possibilities to see
the considerably higher flux than time-averaged UHECR flux from all
sources. For the source number density $n_s={10}^{-5} \,
{\rm{Mpc}}^{-3}$, we expect $E_c \sim 1.4 
\times {10}^{20}$ eV for proton sources. For $n_s={10}^{-4} 
\, {\rm{Mpc}}^{-3}$, $E_c \sim 2.5 
\times {10}^{20}$ eV is expected.

Note that, although we consider the extreme case where each population
has the typical UHECR luminosity respectively, there are not
necessarily clear differences between the two populations. 
For more detailed studies, we should take into account the
distribution of the UHECR luminosity (i.e., luminosity-weighted model).




\clearpage






\begin{thebibliography}{}
\bibitem{Typ1}Note that expressions of
$dn_{\rm{CR}}/d\varepsilon_{\rm{CR}}$ and $n_{\varepsilon, \rm{max}}$ in
Ref. \cite{Mur4} included trivial typos. The expressions given in
this paper is correct. Of course, all the results in Ref. \cite{Mur4} were
calculated using the correct expressions.
\bibitem{Com0}Actually, the adiabatic loss and escape of cosmic rays
should be taken into account carefully, because the cosmic-ray
spectrum could be modified by these effects. Unfortunately,
quantitative estimation is difficult and it is beyond the scope of this paper.
\bibitem{Com2}The dependence on the composition can be seen especially
when $f_{N \gamma}>1$ \cite{Anc1}.
\bibitem{Typ5}Too large baryon loading factors can change a seed
photon spectrum below $\sim$GeV significantly, which may seem 
implausible. Hence, the baryon loading factor could be constrained
from observations of high-energy gamma rays.
However, note that the amount of hadronic gamma rays largely depends on 
$f_{p\gamma}$ when $f_{p \gamma} \lesssim 1$. This implies that the
flux of proton-induced gamma rays is rather sensitive to the lower
photon index $\alpha$. Although there were some theoretical
calculations \cite{Der6,Asa2}, the lower photon index is $\alpha \sim
1.5$ expected in fast cooling synchrotron spectra, rather than the
typical value $\alpha \sim 1$. It can lead to the overestimation of
hadronic gamma rays.
\bibitem{Com3}The intergalactic magnetic field is not necessarily the
magnetic field in the structured region, which is expected to be
relatively strong. In fact, the mean free path of TeV photons is a few
$\times$ Mpc, so that pairs can be generated in the void region whose
magnetic field may be very weak. Note that UHECRs from sources in 
the structured region can be deflected only by the magnetic field in
the structured region, even if we do not consider the magnetic field in
the void region.
\bibitem{Com4}We do not take into account cosmogenic gamma rays 
produced during the UHECR propagation, which can also be expected if 
the intergalactic magnetic field is very weak \cite{Wax8}.
\bibitem{Typ2}More precisely, in order to evaluate the total radiation
energy rate, we have to take into account that the detection flux
threshold is different among detectors.
\end{thebibliography}
\end{document}